\documentclass[referee]{aa} % for a referee version
\usepackage{graphicx}
\usepackage{lscape}
\usepackage{deluxetable}
\usepackage{longtable}
\usepackage{txfonts}
\begin{document}
   \title{Intra-night optical variability of core dominated radio quasars: the role of optical polarization }

   \author{Arti Goyal
          \inst{1,2}
          \and Gopal-Krishna\inst{1} \and Paul J. Wiita\inst{3} \and G. C. Anupama\inst{4} \and  D. K. Sahu\inst{4} 
          \and  R. Sagar\inst{2} \and  S. Joshi\inst{2}
          }

   \institute{National Centre for Radio Astrophysics/TIFR, Pune University Campus, Pune 411 007, India
              \email{arti@ncra.tifr.res.in}
            \and   Aryabhatta Research Institute of Observational Sciences (ARIES), Manora Peak, Naini Tal 263 129, India
            \and   Department of Physics, The College of New Jersey, PO Box 2718, Ewing, NJ 08628-0718, USA 
            \and   Indian Institute of Astrophysics (IIA) Bangalore 560 034, India \\
             }

   \date{Received \today; accepted \today}

% \abstract{}{}{}{}{} 
% 5 {} token are mandatory
 
  \abstract
    { Rapid variations in optical flux are seen in many quasars and all blazars.  The amount of variability
	in different classes of Active Galactic Nuclei has been studied extensively but many questions
	remain unanswered. 
  %We present here evidence that intra-night optical variability (INOV) properties
  %of powerful flat-spectrum radio core-dominated quasars (CDQs) are significantly 
  %dependent on their degree of optical 
  %polarization and relativistic beaming is not a sufficient requirement for strong INOV.    
   }
   {We present the results of a long-term programme to investigate the intra-night optical
   variability (INOV) of powerful flat spectrum radio core-dominated quasars
   (CDQs), with a focus on probing the relationship of INOV to the degree of optical
   polarization.}
   {We observed a %assembled from the literature a 
   sample of 16
   bright CDQs showing strong broad optical emission lines and consisting of both high
   and low optical polarization quasars (HPCDQs and LPCDQs).
   In this first systematic study of its kind, we employed the 104-cm Sampurnanand
   telescope, the 201-cm Himalayan Chandra telescope and the 200-cm IUCAA-Girawali
   Observatory telescope, to carry out {\it R}-band monitoring on a total of 47
   nights. Using the CCD as an N-star photometer to densely monitor each quasar
   for a minimum duration of about 4 hours per night, INOV exceeding $\sim$
   1--2 per cent could be reliably detected.
   Combining these INOV data with those taken from the literature, after ensuring
   conformity with the basic selection criteria we adopted for the 16 CDQs
   monitored by us, we were able to increase the sample size to 21 CDQs
  (12 LPCDQs and 9 HPCDQs) monitored on a total of 73 nights.
   }
   { 
    As the existence of a prominent flat-spectrum radio core signifies that 
    strong relativistic beaming is present in all these CDQs, the definitions 
    of the two sets differ primarily in fractional optical polarization, 
    with the LPCDQs showing a very low median$ P_{op} \simeq$ 0.4 per cent.
    Our study yields an INOV duty cycle
    (DC) of $\sim$28 per cent for the LPCDQs and $\sim 68$ percent for HPCDQs.
    If only strong INOV with fractional amplitude above 3 per cent is considered, the
   corresponding DCs are $\sim$ 7 per cent and $\sim$ 40 per cent, respectively. }
   { 
   From this strong contrast between the two classes of luminous, relativistically beamed
   quasars, it is apparent that relativistic beaming is normally not a
   sufficient condition for strong INOV and a high optical polarization is the other
   necessary condition. Moreover, the correlation is found to persist for many years 
   after the polarization measurements were made.
   Some possible implications of this result are pointed out, particularly in
   the context of the recently detected rapid  $\gamma$-ray variability of blazars.
   }
   \keywords{
    galaxies: active --- galaxies: jets --- polarization --- quasars: general 
               }

   \maketitle
%
%________________________________________________________________

\section{Introduction} 
\label{cdq_intro} 

The occurrence of intranight optical variability (INOV), or  
microvariability, among quasars, particularly their more active subset  
called blazars, is now well documented in the literature (e.g., Miller et 
al.\ 1989; Jang \& Miller 1995, 1997; Romero et al.\ 1997, 2002; 
Gopal-Krishna et al.\ 2003, 2011; Sagar et al.\ 2004, Stalin et al.\ 
2004a,b, 2005; Gupta et al.\ 2005, 2008; Rani et al.\ 2010;  Goyal et al.\ 
2010). Considerable uncertainty persists, however, about the underlying 
physical mechanism and even from a purely observational  perspective
contrasting claims have been made (reviewed, e.g., by Wiita\ 2006; Wagner 
\& Witzel 1995). While some observers find INOV to be more dramatic 
during the optically bright phase of a blazar (e.g., Osterman-Meyer et al.\ 
2009), the opposite has been concluded in another study (Carini 1990).
Moreover, some authors have even reported that INOV is more likely to occur 
when the long-term flux is undergoing a change, 
rather than at some specific flux levels (e.g., Howard et  al.\ 2004; 
Mihov et al.\ 2008). 

In the prior publications under the present long-term programme
(e.g., Gopal-Krishna et al.\ 2003; Stalin et al.\ 2004a, b; 2005; Sagar et al.\ 2004 ), 
an attempt was made to find clues about the INOV phenomenon 
by determining and comparing the INOV characteristics of four major classes 
of powerful active galactic nuclei (AGN). These classes are: `low-frequency-peaked' 
BL Lacs (LBLs, see, e.g., Table 1 of Abdo et al. 2010) whose synchrotron 
emission peaks in the IR/optical range, radio core-dominated quasars (CDQs) 
mostly of the low optical polarization type (LPCDQs), radio lobe-dominated 
quasars (LDQs) and radio-quiet quasars (RQQs).
The study was based on fairly densely sampled intranight $R$-band
differential light curves of duration $\ga$ 4 hours per night for every single AGN
and a minimum of 3 nights for each AGN (totalling 113 nights on a 1-meter telescope),
all processed in a uniform way. 
This study showed that strong INOV (with fractional variability amplitude $\psi > 3$ per cent) is 
exhibited almost exclusively by LBLs and possibly HPCDQs, the high optical 
polarization subset of CDQs, both together termed {\it blazars} 
(e.g., Angel \& Stockmann 1980; Wills et al. 1992; Urry \& Padovani 1995), and that the duty 
cycle of such strong INOV is around 50\% (Gopal-Krishna et al.\ 2003; Stalin 
et al.\ 2004a; Sagar et al.\ 2004; also, Carini et al.\ 2007), very similar 
to the value recently estimated for the subset of blazars detected at TeV energies
(Gopal-Krishna et al.\ 2011). In contrast, the other two classes of radio-loud 
AGN, namely LDQs and LPCDQs, were found to exhibit only low-level INOV and 
that too with a small duty cycle of only around 10-15\%, which is akin to the 
INOV behaviour exhibited by RQQs (Stalin et al.\ 2004a, b; also, Ram\'irez et 
al.\ 2009). 

These findings suggest that radio loudness (even if 
associated with relativistic beaming, as likely in the case of LPCDQs) is not a critical 
factor for the low-level INOV.
% and secondly, that it is not a sufficient condition 
%for strong INOV (which is not exhibited by LDQs). 
Here it may be recalled that
a similarity between the INOV duty cycles of RQQs and core-dominated quasars
had also been noted by de Diego et al.\ (1998). However, an assessment of
their result is rendered difficult due to the fact that many objects in their sample of 17
radio-loud quasars are actually {\it not} core-dominated but, 
instead, have steep radio spectra and are therefore lobe-dominated;
moreover, that study is based on rather sparcely sampled light curves. 
%and followed an uncommon data analysis methodology.

% for LPCDQs (i.e., non-blazar type FSRQs) only low-level INOV has been 
%consistently observed. for the sets of LPCDQs (i.e., non-blazar type FSRQs) 
%monitored in our decade long INOV program (Sect.\ 1) as also in other 
%studies (e.g., Ram{\'i}rez et al.\ 2009). 

A major shortcoming of our afore-mentioned INOV program has been that out 
of the 5 CDQs monitored, only one is a high optical polarization quasar (HPCDQ). 
This precluded a probe into the role of optical polarization in the INOV 
phenomenon. The main purpose of the present study is to rectify this situation, 
by monitoring a set of CDQs which is not only larger in size, but is also a 
balanced mix of HPCDQs showing high optical polarization with 
$P_{op} > 3$ per cent (the canonical benchmark for blazars, Moore \& Stockmann 
1981; Moore \& Stockman 1984; Stockman, Moore \& Angle 1984; Nartallo et al. 
1998; Wills et al. 1992) and their non-blazar 
counterparts, the `low polarization core-dominated quasars' (LPCDQs). It may 
be recalled that even though $P_{op}$ of LPCDQs nearly always remains below 
$\sim 2\%$ (Stockman et al.\ 1984; Schmidt \& Smith 2000), some contribution 
from blazar activity cannot be excluded (e.g., Schmidt \& Smith 2000;
Czerny et al. 2008; Chand et al. 2009). 
A famous example is the nearby LPCDQ 3C 273, a superluminal source whose 
$P_{op}$ always remains below $3\%$ and yet its sensitive photo-polarimetry 
has revealed a `mini-blazar' component (Impey, Malkan \& Tapia 1989; Wills 1989; 
also, Lister \& Smith 2000; Schmidt \& Smith 2000). In rare instances, 
the `mini-blazar' component may undergo a strong flaring, as exemplified by
the quasar 1633$+$382; known to have $P_{op} < 3$ per cent all along, it was 
found in February 1999 to be strongly polarized with $P_{op} = 7.0 \pm 0.5$ 
per cent, confirming its transformation to a bona-fide blazar 
(Lister \& Smith 2000 and references therein). Thus, while in general the 
possibility that $P_{op}$ of a blazar might occasionally dip below 3\% (e.g., 
Moore \& Stockman 1984; Lister \& Smith 2000) should be kept in mind, the 
division at $P_{op} = 3\%$ to discriminate between LPCDQs and HPCDQs, as 
adopted here, remains largely valid and is consistent with many previous 
studies (e.g., Algaba et al. 2011).
%; Smith et al. 2007 and references therein).

Historically, rapid flux variability, high fractional polarization and 
radio-core dominance (i.e., a flat radio spectrum) have all been regarded as 
different facets of blazar activity which, in turn, is believed to be associated 
with a relativistically beamed jet of nonthermal radiation. Starting from the 
discovery of a strong correlation between radio core dominance and $P_{op}$ 
(Impey et al. 1991; Wills et al.\ 1992; also, Lister \& Smith 2000; Fan \& 
Zhang 2003), a flat/inverted radio spectrum of a quasar has often been 
deemed adequate for classifying it as a blazar  (e.g., Maraschi 
\& Tavecchio 2003; Meyer et al.\ 2011). Some authors have even termed FSRQs 
as `strong line blazars' (e.g., Perlman et al.\ 2008), echoing the inference 
reached in Wills et al.\ (1992) that high optical polarization quasars and 
flat-spectrum (core-dominated) quasars are essentially the `same objects'. 
The question specifically examined here is how the rapid optical continuum 
variability (on intranight time scale) relates to the two blazar 
indicators, namely, optical polarization and radio core-dominance. 

In spite of the vast literature exploring the inter-relationships among 
the aforementioned major AGN classes, namely LBLs, HPCDQs, LPCDQs and RQQs,
the picture remains unclear. 
One extreme suggestion bearing on the issue of `radio loudness dichotomy' of 
quasars (e.g., Ivezi\'c et al. 2002) is based on an analogy with the galactic 
micro-quasars. It has been argued that a given quasar becomes radio loud when 
it moves from the `coupled'  to  `flaring' mode of energy production (Nipoti, 
Blundell \& Binney 2005). If true, one will need to revisit the class of 
models in which the weak radio core emission in RQQs is attributed to 
predominantly thermal processes (e.g., Blundell \& Kuncic 2007). Within 
radio-loud quasars, a transition from non-blazar mode (i.e., LPCDQ) to blazar 
mode (HPCDQ), and vice versa, has been quantitatively investigated by Fugmann (1988) 
who estimated that at any epoch nearly two-thirds of flat-spectrum radio quasars 
(FSRQs) exhibit other blazar-like properties (e.g., $P_{op} > 3$ per cent)
(see, also, K{\"u}hr \& Schmidt 1990; Impey \& Tapia 1990). In that case LPCDQs 
and HPCDQs would represent quiescent and active phases of the same FSRQ 
population (see, also, Antonucci \& Ulvestad 1985; Impey et al. 1991). An 
observational hint for such a phase transition comes from the VLBI polarimetric
imaging at 22 and 43 GHz, showing that the magnetic field of the VLBI {\it knots}
in the inner jet is predominently parallel to the inner jet in the case of LPCDQs
(representing the weak shock phase) but orthogonal in HPCDQs (Lister \& Smith 
2000; also, Impey et al.\ 1991). In contrast, for the VLBI {\it cores} of 
LPCDQs and HPCDQs, which manifest the {\it current} activity, no striking 
misalignment dichotomy is found, by considering the difference between their 
radio and optical polarization angles (Algaba, Gabuzda \& Smith 2011). 
Since the typical time scale for the putative phase transition in quasars is
poorly known at present, it is not possible to assess if some of the LPCDQs in 
our sample are in reality HPCDQs that were not observed in an active state. 
We note, however, that in general the putative HPCDQ$\leftrightarrow$LPCDQ transition cannot be 
frequent, in view of the conspicuous correlation observed between high $P_{op}$ 
and {\it long-term} optical variability (e.g., Moore \& Stockmann 1981; 
also Impey et al. 1991; Fan 2005). We shall return to this point in Sect. 5.
A related point to note here is that in many FSRQs a significant contribution 
to the optical continuum can come from the `big blue bump', which is commonly 
understood as quasi-thermal emission from the accretion disc (e.g., Sun \& 
Malkan 1989; Gaskel 2008) (e.g., recall the case of 3C 273 mentioned above). 
This unpolarized thermal emission would dilute the polarized contribution to 
the optical continuum arising from the jet's beamed synchrotron emission (e.g., 
Schmidt \& Smith 2000; Berriman et al. 1990, Giommi et al. 2012), thus diminishing the chance of 
detecting any large INOV associated with the nonthermal relativistic jet. 
Lastly, we note that at the other extreme there are hints that LPCDQs and HPCDQs 
may differ at a more basic level (e.g., Moore \& Stockman 1984; Linford et al. 2011). 
Scarpa \& Falomo (1997) report that LPCDQs have a flatter and 
less smooth optical continuum as well as $\sim 6$ times stronger optical line 
emission {\it intrinsically}, suggesting that their {\it dominant} radiation 
processes might themselves differ.
 
The present work, which is the first systematic study devoted to comparing 
the INOV characteristics of LPCDQs and HPCDQs, is expected to shed light on 
the relationship between these two classes of relativistically beamed radio 
quasars, in particular the relative roles of optical polarization and 
relativistic beaming mechanisms in causing INOV. Our present 
sample consists of 21 flat-spectrum, radio core-dominated quasars (FSRQs/CDQs).
It includes 12 LPCDQs and 9 HPCDQs, each showing strong broad optical emission 
lines. The main difference between the definition of these two sets is in the 
degree of optical polarization (as published in the literature many years ago). 
Out of these quasars, 9 LPCDQs and 7 HPCDQs have been newly monitored by us; 
the INOV data for the remaining 3 LPCDQs and 2 HPCDQ have been taken from the 
literature. For each of these 21 sources, the monitoring duration was $\ga 4$ 
hours (in the R-band) and an INOV detection threshold $\psi \sim 1-2$ per cent 
was reached. Section 2 provides details of our sample selection criteria and 
summarizes the basic properties of our two quasar sets. The observations are 
described in Section 3 and the results in Section 4.   Following 
a brief discussion our conclusions are presented in Section 5.

\section{Sample Selection} 
\label{cdq_sample} 
Since our aim here is to examine the relationship between INOV and the degree 
of optical polarization, we have assembled from the literature (see below)  
two sets of CDQs such that they differ primarily in their optical polarization  
and are similar in other basic properties. Our low polarization sample contains only  
the quasars with $P_{op} < 2\%$ (e.g., Stockman, Moore \& Angel 1984), 
whereas $P_{op} > 3\%$ is the selection criterion adopted for our set of highly polarized 
quasars (see, e.g., Stockman \& Angel 1978; Moore \& Stockman 1981). 
The candidates  shortlisted using the optical polarization data were subjected 
to the following  additional selection criteria, using the data provided in 
V\'eron-Cetty \& V\'eron  (2006): (i) a flat or inverted radio spectrum 
between 1.4 and 4.8 GHz, i.e.,  $\alpha_r > -0.5$, where $S_{\nu} \propto 
\nu^{\alpha_r}$, so as to ensure radio  core-dominance; (ii) $m_B$ $\le$ 18.0 
mag, in order that an INOV detection threshold of $ \psi \sim$1-2 per cent is 
reachable using the 1--2 metre class  telescopes available to us; (iii) 
declination in the range $-$10 to $+$40 deg,  as required for an optimal 
continuous monitoring for at least 5--6 hours with  the telescopes available; 
and (iv) $M_{B}$ $\leq -$23.5 mag, in order to ensure a negligible   
contamination from the host galaxy (e.g., Stalin et al.\ 2004b, Cellone et al.\ 
2007).  
It may be noted that the CDQ/LDQ classification can be epoch dependent,
conceivably due to flux variability of the radio core. We find that this
possibility will have negligible effect on sample definition. To check this
we have computed for each source in our sample the radio spectral index
($\alpha$ between 2.7 and 5 GHz) as published in the quasar catalogue by 
V{\'e}ron \& V{\'e}ron (1996) and, independently from much more recent
measurements reported in Table 1. Both values of $\alpha$ are given in
Table 1. 
Reassuringly, no evidence was found for a change in spectral classification from CDQ 
to LDQ, or vice versa.
 
\subsection{The LPCDQ sample} 
 
This sample of 12 LPCDQs with $P_{op} < 2$ per cent was assembled as follows: 
 
(a) By selecting all 5 LPCDQs in the right ascension range $22^h - 14^h$
    from the optical polarization survey 
    by Wills et al.\ (1992; their Table 1). The LPCDQs are J0741$+$3112,  
    J0842$+$1835, J1229$+$0203, J1357$+$1919 and J2203$+$3145.\\ 
 
(b) We selected the LPCDQ J0005$+$0524 from the UV polarimetry sample of 
    Koratkar et al.\ (1998), which is the only object in their 
    sample of 6 quasars that satisfies all the above criteria.\\ 
 
(c) In order to augment the sample, we included all 3 CDQs from the 
    sample of Sagar et al.\ (2004), for which Wills et al.\ (1992) give $P_{op}  
    < 2$ per cent. These LPCDQs are J0958$+$3224, J1131$+$3114 and J1228$+$3128; 
    J1312$+$3515 was not included as it is a radio-intermediate quasar (Goyal et al.\ 2010). 
   The intranight lightcurves for these 3 LPCDQs are
    taken from  the study by Sagar et al.\ (2004) which belongs to the first 
    part of our INOV programme.\\ 
 
(d) Since the RA range from $23^h$ to $7^h$ still remained sparsely 
    represented, we searched for a few more candidates in this region using the 
    polarization sample of Stockman et al. (1984). In order to keep 
    the numbers manageable, we adopted slightly tighter selection criteria and 
    thus selected only the LPCDQs falling in the declination range of $ -5$  
    to $+10$ deg and having {\it V}-mag brighter than 16.5 (as given in their Table 1). 
    This gave us 6 LPCDQs: J0044$+$0319, J0207$+$0242, J0235$-$0402, J0456+0400,  
    J2346+0930, J2352$-$0105. Out of these, LPCDQs J0044$+$0319 and J0207$+$0242  
    have steep radio spectra ($\alpha_r = -0.67$ and $-0.55$, respectively) while  
    J2352$-$0105 is a known lobe-dominated quasar, again not a CDQ (Stalin  
    et al.\ 2004b). The 3 qualifying LPCDQs (J0235$-$0402, J0456$+$0400 and  
    J2346$+$0930) were included in the sample and monitored by us. Note that 
    although J0235$-$0402 is listed as a steep spectrum object ($\alpha_{r} =   
    -0.62$) in the compendium of V\'eron-Cetty \& V\'eron (2006), it is stated 
    to have a prominent flat spectrum core in the Parkes Half-Jansky sample of 
    flat spectrum sources (Drinkwater et al. 1997), with $\alpha_{2.7}^{5.0} = 
    -0.49$ for the integrated emission.  
 
It may further be noted that all these LPCDQs are bona-fide radio loud quasars, 
each having a radio-loudness parameter (Stocke et al.\ 1992) above 200, 
with the median value for the entire set being $\sim 10^3$. 
It is conceivable that our criterion for selecting LPCDQs,
namely a flat/inverted spectrum around a few gigahertz, also picks
`gigahertz-peaked-spectrum' (GPS) sources which are mostly known
to have low optical polarization (e.g., O'Dea 1998).   A possible
example of a GPS in our sample is the LPCDQ J0741+3112. We note, however, that
the nature of GPS {\it quasars} is still unclear and in several studies
(e.g., Tornianen et al.\ 2005; Tinti et al.\ 2005) their
peaked radio specrum has been attributed to a relativistically beamed
jet, which is akin to HPCDQs, but in stark contrast to GPS galaxies where
the jet is believed to play a negligible role in causing the GPS spectrum 
(see, also, Stanghellini 2003; Bai \& Lee 2005).

\subsection {The HPCDQ sample} 
 
This sample consists of 9 CDQs, all having $P_{op} > 3$ per cent, i.e.,  
well above the maximum value that could normally occur due to dust scattering 
(Impey et al. 1991). 
In this case, the V\'eron-Cetty \& V\'eron (2006) data were used not only 
for applying the aforementioned secondary selection criteria we employed 
for the LPCDQ sample, but also for implementing the primary criterion of a 
high optical polarization $(P_{op} > 3$ per cent). Thus, we shortlisted 
the candidates from the literature (see below) after first ensuring 
that they are labeled as ``HP'' in the compendium of V\'eron-Cetty \& V\'eron (2006). 
The subsequent application of the aforementioned secondary criteria left us 
with 9 quasars (i.e., HPCDQs).  Being highly polarized these 9 flat-spectrum 
radio sources with strong broad emission lines can be termed as bona-fide blazars. 
Details of the selection process are given below:\\ 
 
(a) We selected 7 HPCDQs from the polarization survey of Wills et al. (1992) by
    limiting ourselves to the right ascension range $02^h$ - $15^h$
    and the declination range $-$$10^\circ$ to +40$^\circ$. This yielded 
    the HPCDQs J0239+1637, J0423$-$0120, J0739$+$0136, J1058$+$0133, J1159$+$2914, 
    J1256$-$0547 and J1310+3220.\\ 
     
(b) One HPCDQ, J1218$-$0119, was taken from the first part of our INOV programme   
    (Sagar et al.\ 2004, Stalin et al.\ 2005). The intranight lightcurves were 
    taken from these papers for this source as well as for another two HPCDQs 
    (J0239$+$1637 and  J1310$+$3220) that are part of our set taken from Wills 
    et al.\ (1992), as mentioned above. Note that  these are the only 3 HPCDQs 
    monitored in the first part of our INOV programme. \\ 
 
(c) Lastly, one HPCDQ was taken from the sample of Romero, Cellone \& Combi  
    (1999). They reported V-band intranight monitoring of a sample of southern  
    AGN that contains 4 HPCDQs according to the V\'eron-Cetty \& V\'eron (2006)  
    classification; these are J0538$-$4405, J1147$-$3812, J1246$-$2547, and  
    J1512$-$0906. 
    Since Romero et al.\ (1999) have provided INOV data for just one or two nights for 
    all the sources, these could not be included in the sample straightaway. However,
    J1512$-$0906 is reachable from ARIES; hence, we have included it in the 
    sample and monitored it in the R band for 3 nights.

\subsection{Basic parameters of the two samples} 
 
Table~\ref{tab_cdq_parameter} lists the basic data for our sample. The values  
of extended radio luminosity ($P_{ext}$) and the radio core-dominance parameter  
($f_c$, the ratio of core-to-extended radio luminosities at 5 GHz in the rest 
frame of the source), have been  
determined using the available VLBI measurements at milliarcsec resolution and  
the integrated NVSS flux values at 1.4 GHz, taking a radio spectral index  
of zero for the core ($\alpha_{c} =0$) and $\alpha_{ext} = -0.5$ for the 
extended radio emission. It may be cautioned that the core fluxes of the 
quasars are known to vary (e.g., Savolainen et al. 2002), so the core fraction 
may change with epoch. Since the VLBI observations did not resolve LPCDQ 
J0235$-$0402, we have only computed its total luminosity at 5 GHz using the 
spectral index for the integrated emission (Table ~\ref{tab_cdq_parameter}). 
The absolute blue magnitudes, $M_B$, have been calculated taking 
the total galactic extinction from Schlegel, Finkbeiner \& Davis (1998) and 
assuming an optical  spectral index $\alpha_{op}$ of $-0.7$.  
The concordance cosmological model was assumed, with a Hubble
constant $H_{0} = 70$ km sec$^{-1}$ Mpc$^{-1}$, $\Omega_{m} =0.3$ and
$\Omega_{\Lambda} =0.7$ (Bardelli et al.\ 2009).  
  
\section{Observations} 
 
\subsection{Instruments employed} 
 
The vast majority of these observations was carried out using the 104-cm 
Sampurnanand telescope (ST) located at 
Aryabhatta Research Institute of observational sciencES (ARIES), 
Naini Tal, India. The ST has Ritchey-Chr\'etien (RC) optics with 
a f$/$13 beam (Sagar 1999). The detector was a cryogenically 
cooled 2048 $\times$ 2048 chip mounted at the Cassegrain focus. 
This chip has a readout noise of 5.3 e$^{-}$/pixel and a gain of  
10 e$^{-}$$/$Analog to Digital Unit (ADU) in slow 
readout mode. Each pixel has a dimension of 24 $\mu$m$^{2}$ which corresponds 
to 0.37 arcsec$^{2}$ on the sky, covering a total field of 13$^{\prime}$ $\times$ 
13$^{\prime}$. Our observations were carried out in 2 $\times$ 2 binned mode to 
improve the signal-to-noise ratio. 
The seeing mostly ranged between 
$\sim$1$^{\prime\prime}.5$ to $\sim$3$^{\prime\prime}$, as  
determined using 3 sufficiently bright stars on the CCD frame;  
plots of the seeing are provided for all of the nights 
in the bottom panels of Figs.\ 1 and 2  (see Sect.\ 4.1). 

We also used the 201-cm Himalayan Chandra Telescope 
(HCT) at the Indian Astronomical Observatory (IAO), located in Hanle, India. 
This telescope is also of the RC design but has a f$/$9 beam at the 
Cassegrain focus\footnote{http://www.iiap.res.in/$\sim$iao}. 
The detector was a cryogenically cooled 2048 $\times$ 4096 chip, of which the 
central 2048 $\times$ 2048 pixels were used. The pixel size is 15 $\mu$m$^{2}$, 
so that the image scale of 0.29 arcsec$/$pixel 
covers an area of 10$^{\prime}$ $\times$ 10${^\prime}$ on the sky.  
The readout noise of CCD is 4.87 e$^{-}$/pixel and the gain is 1.22  
e$^{-}$$/$ADU. The CCD was used in an unbinned mode.  
The seeing ranged mostly between $\sim$1$^{\prime\prime}$ to  
$\sim$2$^{\prime\prime}.5$. 

Lastly, a few nights of blazar monitoring data were obtained using the 200-cm 
IUCAA Girawali
Observatory (IGO) telescope located near Pune, India. It has an RC design
with a f$/$10 beam at the Cassegrain focus\footnote
{http://www.iucaa.ernet.in/\%7Eitp/igoweb/igo$_{-}$tele$_{-}$and$_{-}$inst.htm}.
The detector was a cryogenically cooled 2110$\times$2048 chip mounted
at the Cassegrain focus. The pixel size is 15 $\mu$m$^{2}$ so that the
image scale of 0.27 arcsec$/$pixel covers an area of 10$^{\prime}$
$\times$ 10${^\prime}$ on the sky. The readout noise
of this CCD is 4.0 e$^{-}$/pixel and the gain is 1.5 e$^{-}$$/$ADU.
The CCD was used in an unbinned mode. The seeing ranged 
between $\sim$1$^{\prime\prime}$.0 and $\sim$2$^{\prime\prime}$.5.

All the observations were made using {\it R} filters, as the CCD  
responses is maximum in this band. 
The exposure time was typically between 12 to 30 minutes for the ARIES observations  
and ranged between 3 to 6 minutes for the observations from IAO and IGO, depending 
on the 
brightness of the source, the phase of the moon and the sky transparency on
that night. The field positioning was adjusted so as to also have  
within the CCD frame at least 2--3 comparison stars.  
For all telescopes  bias frames were taken intermittently, and twilight sky  
flats were also obtained. 
 
\subsection{Data reduction} 
All pre-processing of the images (bias subtraction, flat-fielding and cosmic-ray 
removal) was done by applying standard procedures in the 
{\textsc IRAF} \footnote{\textsc {Image Reduction and Analysis Facility (http://iraf.noao.edu/) }} 
and {\textsc MIDAS}\footnote{\textsc {Munich Image and Data Analysis System http://www.eso.org/sci/data-processing/software/esomidas// }} 
software packages. The instrumental magnitudes of the target AGN (quasars) and the stars in  
the image frames were determined by aperture photometry, using  
DAOPHOT \textrm{II}\footnote{\textsc {Dominion 
Astrophysical Observatory Photometry} software} (Stetson 1987). 
The magnitude of the target AGN was measured relative to the nearby apparently steady comparison 
stars present on the same CCD frame (Table ~\ref{tab_cdq_comp}). In this way Differential Light 
Curves (DLCs) of each AGN were derived relative to 3 comparison stars designated as S1, S2, S3. 
These comparison stars are within about a magnitude
of the target AGN, this precaution being important for minimizing the possibility of
spurious INOV detection (e.g., Cellone et al.\ 2007).
In our study the {\it B-R} colours of quasars and the comparison stars are 
often quite different (Table 2).  However it is shown by Carini et al.\ (1992) and Stalin et al.\ (2004a) 
that such colour differences do not yield a significant amount of spurious INOV due to the
different second-order extinction coefficients of the quasar and the comparison stars as 
they are observed through varying airmass 
during the course of monitoring.   For the airmass range between 1 and 2 the {\it B-R} colour 
difference between the quasar and the comparison star as high as 1.9 causes negligible errors. 

For each night, an optimum aperture radius for photometry 
was selected on the basis of the observed dispersions in the star-star DLCs that were found 
for different aperture radii starting from the median seeing (FWHM)  
value on that night to 4 times that value. We selected the appropriate aperture for each night 
as the one that provided the minimum dispersion 
for the steadiest DLC found among all pairs of the comparison stars 
(e.g., Stalin et al.\ 2004a). Typically, the selected aperture 
radius was $\sim$4$^{\prime\prime}$ and the seeing was found to be $\sim$2$^{\prime\prime}$. 
 
\section{Results}  
\label{results}
\subsection{Differential Light Curves (DLCs)} 
\label{dlcs_tables}
The intranight DLCs for the LPCDQs and HPCDQs observed in our monitoring
 programme are  shown in Figures\ \ref{fig:1} and \ref{fig:2} respectively, 
while the corresponding DLCs  showing their long-term optical variability 
(LTOV) are displayed in Figs.\ \ref{fig:3} and \ref{fig:4}. 
Tables \ref{tab_inov_lpcdq} and  ~\ref{tab_inov_hpcdq} summarize the results 
of the INOV observations of our sets of LPCDQs and HPCDQs made by us and 
augmented with those taken from the literature (Sect.\ 2).  

\subsection{Estimation of the parameter $\eta$}
\label{eta_comp}
It has been found in several published studies that the photometric errors 
returned by the $\sc APPHOT$\footnote{Photometry package in \sc IRAF } package 
are systematically too low such that the rms error for each datapoint is
underestimated by a factor $\eta$, found to range between 1.30 and 1.75 
(Gopal-Krishna et al. 1995; Garcia et al. 1999; Stalin et al. 2004a; Bachev
et al.\ 2005).
To verify and quantify this factor for the present set of observations and the version of the
$\sc APPHOT$ used here, we have made a fresh estimate of $\eta$ as follows.
Out of the 3 star-star DLCs available for each night (using the 3 comparison 
stars monitored), we first selected the steadiest star$-$star DLC. Thus, for
our entire dataset (73 nights) we get 73 `steady' DLCs, whose stars appear
to have not varied on the corresponding night. For each selected DLC with $N_p$ data points, we 
then computed the $\chi^2$ corresponding to its number of degrees of freedom ($\nu$ =
$N_p$ - 1). In Fig. \ref{fig:5}, we plot for each night, the computed $\chi^{2}$ 
value together with its corresponding {\it expectation values} of $\chi^{2}$ 
at $p=0.5$ which corresponds to 50 per cent probability. 
It is seen that for most of the `steady' star-star DLCs the calculated 
$\chi^2$ values lie above their {\it expectation values} when no correction 
factor is applied to the photometric errors (i.e, $\eta$=1, top diagram). 
However, when a correction factor of $\eta = 1.5$, is applied
to all the data points, the computed $\chi^{2}$ values for the 73 nights are found 
to be evenly dsitributed about the solid curve showing the expectation values,
as is indeed expected for the median estimator of the distribution (bottom diagram). 
We therefore adopt $\eta=$1.5, for scaling up the $\sc IRAF$ 
photometric rms errors.

\subsection{ Peak-to-peak INOV amplitude ($\psi$)}

The peak-to-peak INOV amplitude is calculated using the definition of Romero,  
Cellone \& Combi (1999) 
\begin{equation} 
\psi= \sqrt{({D_{max}}-{D_{min}})^2-2\sigma^2} 
\end{equation} 
with  $D_{min,max}$ = minimum (maximum) in the AGN differential light curve, 
and $\sigma^2$= $\eta^2$$\langle\sigma^2_{err}\rangle$. 
where, $\eta$ =1.5.
% for our analysis, see above).

\subsection{INOV detection; $F$-statistics}
\label{ftest_analysis}

Hitherto the criterion most commonly used in the literature for checking 
the presence of INOV is based on the so-called `$C$-statistic', which is 
defined as the ratio of standard deviations of the `QSO-star' DLC and the 
corresponding `star-star' DLC (e.g., Jang \& Miller 1997; Romero et al.\ 
1999; Stalin et al.\ 2004, 2005; Xie et al.\ 2004; Carini et al.\ 2007; 
Gupta et al.\ 2008; Goyal et al.\ 2010).
Recently, de Diego (2010) has emphasized that the usual definition of 
$C$ is not a proper statistic, as it is based on the ratio of standard 
deviations which (unlike variance) are not lineal statistical operators.
They argue that the critical values for the $C-$test are wrongly 
established, being much larger (i.e., more conservative) than those for the
$F-$test which is based on the ratio of variances.
In addition, the commonly employed test based on the $C$-statistic ignores 
the number of degrees of freedom in the observation, which too is inappropriate.
A version of the $C$-statistic that properly incorporates degrees of 
freedom can be devised (Villforth, Koekemoer \& Grogin 2010), but has not 
yet been used in INOV studies. Therefore, in this work we shall employ
the $F$-statistics to quantify INOV detection which is defined as follows 
(Villforth et al. 2010): 
\begin{equation} 
\label{f_def}
F = \frac{observed \hspace{0.5pc} variance}{expected \hspace{0.5pc} variance} = \frac{var_{observed}}{var_{expected}}
\end{equation}
 where $var_{observed}$ is the variance of the flux measurements in a DLC and $var_{expected}$ 
 is the mean of the squares of flux error estimates.

In computing the $F$-value we first examined the `star-star' DLCs derived 
from (typically 3) comparison stars monitored along with the quasar
in the same session (Figures \ref{fig:1} \& \ref{fig:2}), in order to 
select the steadiest DLC out of them. The corresponding two stars are 
designated as CS1 and CS2 (they are not necessarily the stars labelled 
as S1 and S2 in the figures 1 \& 2), with the convention that CS1 is
better matched to the quasar in R-magnitude, compared to CS2. 
After adjusting for the underestimation of the measurement  errors (Sect. ~\ref{eta_comp}) 
by setting $\eta =$ 1.5, $F$-values can be written as, 

\begin{eqnarray}\nonumber 
F_{CS1} = \frac{Var(Q-CS1)}{ \eta^2 \langle \sigma_{Q-CS1}^2 \rangle}, 
F_{CS2} = \frac{Var(Q-CS2)}{ \eta^2 \langle \sigma_{Q-CS2}^2 \rangle}, \\ 
&& \hspace{-17.2pc}  
F_{CS1-CS2} = \frac{Var(CS1-CS2)}{ \eta^2 \langle \sigma_{CS1-CS2}^2 \rangle} 
\end{eqnarray} 
where $Var(Q - CS1)$, $Var(Q - CS2)$ and $Var(CS1 - CS2)$ are the 
variances of the `quasar-CS1', `quasar-CS2' and `CS1-CS2' DLCs and 
$\langle \sigma_{Q-CS1}^2 \rangle$, 
$\langle \sigma_{Q-CS2}^2 \rangle$ and $\langle \sigma_{CS1-CS2}^2 \rangle$ 
are the mean square (formal) rms errors of the individual data 
points in the `quasar-CS1', `quasar-CS2' and `CS1-CS2' DLCs, 
respectively.  

In this way, the  $F$-value was computed for each DLC and compared with
the critical $F$-value, $F_{\nu}^{\alpha}$, where $\alpha$ is the
significance level set by us for the test and $\nu$ (= $N{_p}-1$) is the degree of 
freedom for the DLC. The smaller the value of $\alpha$, the more unlikely is the variation 
to occur by chance. For the present study, we have used two significance levels,
$\alpha$ = 0.01 and 0.05, corresponding to confidence levels of $p >$ 99 per cent 
and $p >$ 95 per cent, respectively.    
Thus, in order to claim a genuine INOV detection, i.e., assigning a designation `variable'
designation (V), we stipulate that the computed $F$-value is above the critical 
$F$-value corresponding to $p > 0.99$. 
 A `possible variable' (PV) designation was assigned when the confidence
level for the DLC was found to be in the range $0.95 < p \le 0.99$, 
while a `non-variable' (N) designation was assigned if $p \le 0.95$.
Tables \ref{tab_inov_lpcdq} and  \ref{tab_inov_hpcdq} summarize the INOV results 
for our sets of LPCDQs and HPCDQs, both the ones monitored by us and those
for which we have taken the DLCs from the literature (Sect.\ 2). 
We have carried out the $F$-test independently for the DLCs of each quasar,
drawn relative to CS1 and CS2, yielding two estimates of the INOV duty
cycle (Sect. \ref{inov_cdq_results}) for the LPCDQ set and also for 
the HPCDQ set (Table \ref{dc_stat}). Good
agreement between the two estimates of duty cycle is found, despite the different
levels of brightness mismatches of the quasar from the two chosen
comparison stars (Tables  \ref{tab_inov_lpcdq} and \ref{tab_inov_hpcdq}). 
This provides a {\it post facto} validation of our assumption that the $F$-test 
is not unacceptably sensitive to the typical rms errors on individual data 
points being slightly different for the two DLCs involved in the $F$-test 
for each quasar, namely, `Q - CS1' and `Q - CS2'.   It
needs to be mentioned here that care has been taken that 
the comparison stars are nearly always within 1-mag of the respective 
quasars. (For the LPCDQ set, the median magnitude mismatch is 0.3-mag for
CS1 and 0.8-mag for CS2 and the corresponsing values for the HPCDQ set
are 0.9-mag and 1.4-mag, respectively).
%*** MOVED TO NEXT PARA While quoting the DC estimates for our sets of LPCDQs and HPCDQs (Sect. 4.5),
%we take the average of the two estimates of DC arrived at by using CS1
%and CS2 (see Sect. \ref{inov_cdq_results}). 

It is seen that for a total 11 out of 73 nights, the quasar variability status 
inferred from the DLC using one comparison star (CS1) differs from that 
found using the DLC using the other comparison star (CS2). A possible 
explanation is that one of the stars may have varied.
Since such putative low-level INOV of the comparison star would remain unnoticed by eye and 
hence we have no justification to prefer one comparison star over the 
other (in terms of steadiness), we list in Table \ref{dc_stat} the 
estimates of INOV duty cycle (DC) for each quasar using both comparison
stars, CS1 and CS2 (chosen because their DLC appeared to be 
the steadiest). While quoting the DC estimates for our sets of LPCDQs and HPCDQs in Sect.\ 4.5,
we take the average of the two estimates of DC arrived at by using CS1
and CS2.% (see Sect. \ref{inov_cdq_results}). *** REDUNDANT***

Here it may be recalled that the $F$-test provides less statistical power 
(i.e., more non-detections of actually variable sources) than an alternative 
like the ``analysis of variance", or ANOVA, which tests for differences 
between the mean values, instead of the contrast between the variances (e.g., 
de Diego 2010). However, the relatively long exposures required in our 
measurements means that many of our light curves had fewer than 30 data 
points, precluding us from applying the ANOVA test with sufficient power.

\subsection{The computation of INOV duty cycle (DC)}  
\label{inov_cdq_results} 
The INOV duty cycle was computed following the definition of  
Romero et al.\ (1999) (see, also, Stalin et al.\ 2004a): 
\begin{equation} 
DC = 100\frac{\sum_{i=1}^n N_i(1/\Delta t_i)}{\sum_{i=1}^n (1/\Delta t_i)} {\rm per cent} 
\label{eqno1} 
\end{equation} 
where $\Delta t_i = \Delta t_{i,obs}(1+z)^{-1}$ is duration of the 
monitoring session of a source on the $i^{th}$ night, corrected for 
its cosmological redshift, $z$. Note that since for a given source the 
monitoring durations on different nights were not always equal, the 
computation of DC has been weighted by the actual monitoring duration 
$\Delta t_i$ on the $i^{th}$ night. $N_i$ was set equal  to 1 if INOV 
was detected, otherwise $N_i$ = 0.

Employing the $F$-statistics the computed INOV DCs are: 28 per cent for LPCDQs 
(45 per cent if the `PV' cases are included) based on 44 nights' monitoring 
(Table~\ref{tab_inov_lpcdq}); and 68 per cent (70 per cent if one `PV' 
case is included) for the HPCDQs based on 29 nights' data 
(Table~\ref{tab_inov_hpcdq}). If only the nights showing 
$\psi > 3$ per cent are considered (all of which, clearly, belong to the `V' 
category), the derived DCs are 7 and 40 per cent for LPCDQs and HPCDQs, 
respectively.

At $p = 0.99$, the expected value of  false positives for our data sets 
of LPCDQs (44 nights) and HPCDQs (29 nights) are, 0.44 and 0.29, respectively. 
Thus, in both cases, we expect no more than $\sim$ 1 DLC to be falsely 
classified as variable. 
Similarly, at $p = 0.95$, the expected value of false positives for our 
two data sets of LPCDQs and HPCDQs are $<3$ and $<2$, respectively. 

In order to ensure a consistent analysis and the check on the error estimates, 
we have also estimated the rate of false positives using actual data, namely 
our data sets of LPCDQs and HPCDQs. To do this, we have performed the F-test 
analysis on our set of 73 `steady' star-star DLCs based on the same comparison stars 
that were used to generate the `quasar-star' DLCs we used for computing the DCs. 
. The results are given in 
Tables \ref{tab_inov_lpcdq} \& \ref{tab_inov_hpcdq}. 
This also provides the `sanity check' on our error estimation as returned 
by $\sc APPHOT/IRAF$. From the LPCDQ data set, 2 out of 44 star-star DLCs 
are found designated as `clear variable', while the number for the HPCDQ 
data set is found to be 1 out of 29 star-star DLCs. The good agreement between the {\it expected}  
and {\it observed} rates of false positives for our LPCDQ and HPCDQ data sets validates 
our analysis procedure. 

\subsection{Notes on individual sources } 
\label{inov_notes}
Below we give brief comments on the variability characteristics of some of the  
quasars in our sample. 
 
\begin{itemize} 
\item{{\bf LPCDQ J0741$+$3112:} This CDQ was monitored by us on 4 nights and 
was found to vary only on 21 Jan.\ 2006 and 22 Dec. 2006. It showed  
a very clear, almost sinusoidal light curve with $\psi = 4.9$ per cent.  
Seeing remained stable at $2^{\prime\prime}$ throughout the monitoring 
period (bottom panel; Fig.~\ref{fig:1}).}  

\item{{\bf LPCDQ J1229$+$0203:} Known to be harbouring a mini-blazar
(e.g., Impey, Malkan \& Tapia 1989), this well known CDQ, 3C 273, 
showed INOV on 2 out of the 3 nights it was monitored by us (Fig.~\ref{fig:1}).}

\item{{\bf LPCDQ J1357$+$1919:} This CDQ has been extensively monitored 
in our programme on a total of 8 nights.  
On 28 Mar. 2006, it showed a striking INOV pattern, clearly fading by 
$\sim 2$ per cent  
during the first 2 hours of the monitoring, followed by a steady level 
for the next 1.5 hours and finally a brightening by $\sim 2$ per cent 
in the final 1.5 hours (Fig.~\ref{fig:1}).} 

\item{{\bf HPCDQ J0238$+$1637:} This CDQ has been known for its nearly 100 
per cent INOV duty cycle (e.g., Gopal-Krishna et al.\ 2011), the present data
conform to this (Fig.~\ref{fig:2}). In addition to our single night's 
observations, this CDQ had earlier been monitored in {\it R}-band by Sagar et
al.\ (2004) on 3 nights, and on each occasion INOV was confirmed, with
$\psi$ ranging between 5 to 20 per cent (Table ~\ref{tab_inov_hpcdq}). 
Likewise, Romero et al.\ (2002) found it to vary on each night they 
monitored it, with $\psi$ in the range 7--44 per cent 
(Table ~\ref{tab_inov_hpcdq}).}

\item{{\bf HPCDQ J1159$+$2914:} This CDQ is an OVV quasar (Sitko, Schmidt and Stein 1985). 
We monitored it on 3 consecutive nights and it showed INOV on each night, 
 with $\psi$ exceeding about 5, 9 and 16 per cent (Table ~\ref{tab_inov_hpcdq}; Fig. ~\ref{fig:2}). 
Although the mean brightness of the CDQ remained unchanged between the first 2 nights 
(i.e., 31 Mar. 2012 and 01 Apr. 2012) later it showed strong {\it inter night} 
variability as it brightened by $\sim$ 0.5 magnitude 
between 01 Apr. 2012 and 02 Apr. 2012. }

\item{{\bf HPCDQ J1256$-$0547:} This famous CDQ, 3C 279, is known to have a 
high and variable polarization and was the first flat-spectrum quasar to be 
detected above 100 GeV (Albert et al.\ 2008).  It showed INOV on all the 3 nights 
we monitored it, with $\psi$ values of 4, 10  and 22 per cent 
(Table ~\ref{tab_inov_hpcdq}; Fig. ~\ref{fig:2}).} 
\end{itemize} 
 
\section{Discussion and Conclusions} 
\label{discussion}
 
In the present study we have made a quantitative comparison of the INOV  
characteristics of two sets of bright radio core-dominated quasars,
both showing strong broad optical emission lines but differring markedly in 
fractional optical polarization, $P_{op}$. To illustrate this we display 
in Fig. \ref{cdq_dis_properties} the distributions of $P_{op}$ and five other basic 
parameters for our sets of LPCDQs and HPCDQs. The parameters are: redshift ($z$); 
blue absolute magnitude, ($M_{B}$); radio spectral index ($\alpha_r$); 
radio core-fraction ($f_c$), which is a well known orientation indicator
because the extended radio lobe flux density is essentially independent of 
orientation, while the core flux density is Doppler boosted when 
the radio source axis is oriented near the line-of-sight (e.g., Kapahi \& 
Saikia 1982; Orr \& Browne 1982; Morisawa \& Takahara 1987); 
and luminosity of the extended radio emission ($P_{ext}$) at 5 GHz, which is a 
measure of the jet's intrinsic power (e.g., Willott et al.\ 1999; Punsly 2005).  
Application of the two-sample Kolmogorov-Smirnov test shows that the null
hypothesis that our sets of LPCDQs and HPCDQ belong to the same parent 
population cannot be rejected for the parameters $z$, $M_{B}$, $\alpha_r$, 
$f_c$ and $P_{ext}$ (Table ~\ref{ks-test_cdq}), whereas the hypothesis 
that they are drawn from the same distribution of $P_{op}$, can be rejected 
at high confidence ($>$ 99.9 per cent). Thus, $P_{op}$ is the key discriminator 
between our sets of LPCDQs and HPCDQs.
%{\bf Here it may be relevant to point out that the optical flux of HPCDQs may
%have a significant, sometimes even dominant synchrotron component contributed by
%the relativistic jet. In that event, our HPCDQ set would be systematically weaker
%intrinsically compared to the LPCDQ set. Unfortunately, 
%the results presented here are subject to the caveat since
%it is not possible at present to quantify and account for the jet contribution reliably.
%Nonetheless, there has been a mild evidence that the LPCDQs may have significant optical
%synchrotron component (e.g., Whiting et al. 2001) but on the whole it may be swamped by the
%thermal emission from the accretion disk (e.g., Giommi et al. 2012) resulting in flatter 
%optical spectra for these sources (Moore \& Stockmann 1981; 1984). Also, the jets of LPCDQ and HPCDQ may also be
%intrinsically different, for instance, due to the strong magnetic field structure (Lister \& smith
%2000). Conceivably, such a scenario may partly account for the lower
%polarization of LPCDQs.}

Here it may be relevant to point out that the optical flux of HPCDQs 
may have a significant, even dominant, synchrotron component
contributed by the relativistic jet. In that event, our HPCDQ set 
would be systematically weaker {\it intrinsically} compared to the LPCDQ set,
since they are of similar absolute optical magnitudes. 
Unfortunately, it is not possible at present to quantify and subtract out the  
jet's contribution reliably. Nonetheless, even if any such a bias
is significant for our datasets, that would probably mean that the 
central black holes in our LPCDQs are, on average, more massive than 
those present in our HPCDQ set. Unfortunately, there is at present 
no knowledge about the dependence of INOV on the mass of the central 
black hole, although significant information does exist concerning the 
long-term optical variability (LTOV, on year-like time scales in the rest frame). 
Using large samples of SDSS quasars it has been found that the quasars 
containing more massive central black holes tend to exhibit stronger 
long-term optical variability (Wold, Brotherton \& Shang 2007; 
Bauer et al.\ 2009). 
Thus, at least on the basis of the observed trend 
in the quasar LTOV, which correlates
positively with both optical polarization (Sect. 1) and central black hole mass,
there is little reason to 
suspect that the stronger INOV found here for the HPCDQ set, in comparison 
to the LPCDQ set, results from of the latter being optically more
luminous and hence probably containing more massive central 
black holes.

Our choice of $F$-statistic in the present study (Sect. \ref{ftest_analysis}) 
precludes us from making an exact comparison of the present results with 
those available in the literature (which are mostly based on the 
$C$-statistic, Sect. 1). Our main finding is that even though relativistically
Doppler boosted (radio) jets are prominent in all 12 LPCDQs, the duty cycle 
for strong INOV (DC $\sim 7$ per cent for $\psi > 3\%$) is much smaller than 
that found for their high polarization counterparts, namely the 9 HPCDQs 
(DC $\sim 40$ per cent for $\psi > 3\%$) (Sect.~\ref{inov_cdq_results}). Further, the
result (Table \ref{tab_inov_hpcdq}) that INOV amplitudes above $3\%$ are almost exclusively 
observed for HPCDQs and only rarely seen in LPCDQs (despite their being 
strongly beamed too), makes HPCDQs closely resemble LBLs in their INOV behaviour 
(see Gopal-Krishna et al.\ 2003; 2011; Stalin et al.\ 2004a,b). 
The distributions of $\psi$ for our sets of 12 LPCDQs (44 DLCs; 
Table ~\ref{tab_inov_lpcdq}) and 9 HPCDQs 
(29 DLCs; Table ~\ref{tab_inov_hpcdq}) are compared in Figure 7. It needs to 
be clarified that the intra night monitoring durations are very similar for these sets 
of LPCDQ and HPCDQ, the median values being 5.7 and 5.8 hours, respectively. 
Such a matching is desirable in view of the fact that INOV detection probability is 
at least moderately sensitive to the monitoring duration (e.g., Romero et 
al.\ 2002; Carini et al.\ 2007). A two sample Kolmogorov-Smirnov (K-S) test 
performed on these $\psi$ distributions rejects the null hypothesis that the 
two are drawn from the same parent population, giving it a probability of 
only $3\times10^{-4}$. This statistical comparison confirms that HPCDQs are 
much more prone to display INOV than their weakly polarized counterparts, LPCDQs. 
In stark contrast 
to the HPCDQ set, $\psi$ was found to exceed 4 per cent level only once out 
of the 44 nights of LPCDQ monitoring by us. This occurred for the LPCDQ 
J0741$+$3112 which attained $\psi = 4.9$ per cent on 21 January 2006 
(Table 2). Interestingly, its light-curve on that night showed an 
extraordinary, almost {\it sinusoidal} pattern (Fig. \ref{fig:1}), similar to the 
rare events recorded earlier for the archetypal intra-day variable blazar 
S5 0716$+$714 on the nights of 1 January 2004 (Wu et al.\ 2005) and 1 April 
2008 (Stalin et al.\ 2009). We, therefore, consider the LPCDQ J0741$+$3112 
to be a good candidate where a transition from LPCDQ to HPCDQ phase might 
have occured, as reported for the quasar 1633$+$382 (Sect.\ 1). Hence, 
optical polarimetric monitoring of J0741$+$3112 would be particularly interesting.

The present observations also provide information on long-term optical 
variability (LTOV) on month-like or longer timescales 
(Figs.~\ref{fig:3} \& \ref{fig:4}), we find such variability to be common 
among both LPCDQs and HPCDQs, with
amplitudes approaching 0.1-mag level in the R-band. This result is in 
accord with the findings of Webb \& Malkan (2000) for more common types of 
AGN; for roughly half the AGNs they found optical variability amplitudes of 
0.1 -- 0.2 mag (rms) on month-like time scales. Since the total time span 
covered in our observations differ vastly from source to source, these data 
do not permit a quantitative comparison of the LTOV of the HPCDQs and LPCDQs 
monitored.

In summary, the point emerging from the present study is 
that for strong INOV, optical polarization is a key requirement even when
a strongly beamed synchrotron radio jet is observed (see, Sect. 5). This echoes the 
well known close connection between the optical polarization of quasars
and their long-term optical variability (e.g., Moore \& Stockman 1984; 
Impey et al. 1991). In other words, just as the INOV amplitude and duty 
cycle for powerful AGNs are not automatically bolstered due to radio 
loudness, as already inferred in the first part of our programme from the 
similarities of the INOV levels found for LDQs, LPCDQs and RQQs (see Stalin 
et al.\ 2004b, Stalin et al.\ 2005; also, Ram{\'i}rez et al.\ 2009; Sect. 1), 
the present study provides a strong hint that relativistic beaming (as 
indicated by the radio core dominance) is normally not a sufficient 
condition for the occurence of strong INOV, unless it is accompanied by a 
strong optical polarization. Furthermore, this trend exists even if the polarization 
was measured in relatively distant past (see below).

Thus, even though the polarized optical flux is widely regarded as a 
manifestation of relativistically beamed nonthermal emission (e.g., 
Malkan \& Moore 1986; Impey et al. 1991), the physical connection of 
optical polarization to INOV appears to supercede the link between INOV 
and relativistic beaming. This is evident from the much more modest INOV 
found for LPCDQs, even though they are core-dominated like the HPCDQs 
and hence also possess relativistically beamed jets. Now, it is 
conceivable that the jets are so curved that their inner, optically  
radiating, beamed segments are misdirected from us (evidence for bending 
between the sub-parsec and parsec scales does exist for blazars, e.g., Lobanov \& 
Zensus 1999; Readhead et al.\ 1983). However, this explanation is unlikely 
to account for the persistent lack of strong INOV among LPCDQs, firstly 
since jet bending on sub-parsec scale is much milder for LPCDQs (Impey 
et al. 1991) and, secondly because it is known to vary on month/year-like 
time scales (e.g., Britzen et al.\ 2010 and references therein), whereas 
the optical polarization measurements used for selecting our LPCDQ set 
were carried out more than a decade ago (Sect.~\ref{cdq_sample}). This 
then suggests that the propensity of a given radio core-dominated quasar 
to exhibit strong INOV is of a fairly stable nature and it correlates 
rather tightly with optical polarization class. This inference may appear 
to run counter to the notion that FSRQs keep switching between high- and 
low-polarization states (HPCDQ $\leftrightarrow$ LPCDQ; Sect. 1), in case the typical
time scale for such transitions is much shorter than the decade$-$like 
time interval between their optical polarimetric classification and their
INOV observations reported here. Conceivably, such polarimetric phase 
transitions could occur on fairly short, say, year-like time scales that 
characterise successive ejections of blobs of synchrotron plasma (VLBI 
knots) out of the central engine (e.g., Aller, Aller \& Hughes 2006; Bell 
\& Comeau 2010; Hovatta et al.\ 2007; Le{\'o}n-Taveres et al.\ 2010; also,
Impey \& Tapia 1990). 
However, were this indeed the case, then during the decade long time 
interval elapsed since the original optical polarimetry, the $P_{op}$ 
distribution within the sets of LPCDQs and HPCDQs would have gotten
substantially randomized by the time their INOV observations took place.
Consequently, little difference should have been found between the 
INOV duty cycles for the LPCDQ and HPCDQ sets, in clear contradiction to the 
present result. For a more direct check on this, a fresh round of optical 
polarimetry is encouraged for the sets of LPCDQs and HPCDQs (which are 
fairly bright, Table \ref{tab_cdq_parameter}), particularly for the 
two LPCDQs which have exhibited unusually strong INOV ($\psi > 3$ per 
cent) during our monitoring (Sect. \ref{inov_notes}).
 
Recent radio VLBI and optical (and sometimes even X-ray) monitoring 
observations of a few blazars have provided useful insight into the likely 
physics behind the flaring and polarization of their emission. 
According to an emerging picture (e.g., D'Archangelo et al.\ 2007; Jorstad 
et al.\ 2007; Marscher et al.\ 2008; Arshakian et al.\ 2010; Le{\'o}n-Taveres 
et al.\ 2010), much of the polarized optical and radio synchrotron flux and 
its flaring arise as the successive energetic disturbances emanating from the 
central engine and then traversing the helical magnetic field along the jet's 
initial acceleration/collimation zone, cross through a standing shock in the 
jet. Such standing shocks typically form at a projected distance of a few 
parsecs from the central engine, where particle acceleration takes place 
%(e.g., Mandal \& Chakrabarti 2008; Becker, Das \& Le 2008) 
and the inflowing synchrotron plasma is locally compressed. In this scenario, 
the magnetic field near the end of the jet's acceleration zone (which may 
extend from the central engine up to $\sim 10^4$ times the gravitational 
radius of the supermassive black hole, e.g., Vlahakis \& K\"{o}nigl 2004; 
Meier \& Nakamura 2006) is predominantly longitudinal to the jet, probably 
due to velocity shear (e.g., Jorstad et al. 2007). However, a build-up of 
turbulence in this region, e.g., due to weakening of the collimating helical 
magnetic field (e.g., Arshakian et al.\ 2010), or some externally induced 
disturbance (see below), can locally generate a substantial {\it transverse} 
component of magnetic field in the flow. As this turbulent jet plasma
passes through the standing shock downstream, not only will particle 
acceleration and the plasma compression take place, boosting the multi-band 
synchrotron output, but the same compression would also amplify any transverse component of the pre-shock 
magnetic field (e.g., caused due to the turbulence, as mentioned above), 
giving rise to an enhanced polarization signal (e.g., Hughes et al. 1985; 
Marscher \& Gear 1985; Laing 1980). If now the postulated zone of turbulence 
in the jet, just upstream of the standing shock, is identified as the site 
where the bulk of INOV arises, then the scenario sketched here may provide 
a plausible explanation for the close link of INOV to optical polarization 
underscored in this study. Conversely, if a strong confining helical field 
persists in LPCDQs, this would tend to subdue the growth of turbulence in the jet plasma, 
leading to both a weak INOV and a milder build-up of the transverse component of
magnetic field in that region. The latter would then result in only a 
modest field amplification as the jet plasma undergoes compression while 
crossing the first (transverse) standing shock.  
%For longitudinal magnetic field in the inner jets of LPCDQs, see Sect. 6.4.1
%of Lister \& Smith (2000) 
An observational constraint which this simple picture must satisfy is that 
the postulated zone of turbulence upstream of the standing shock in blazar 
jets must be a fairly long lasting feature, for consistency with the 
observed persistence of strong INOV we find for HPCDQs vis {\`a} vis LPCDQs, even 
a decade after their optical polarimetry was carried out and the 
HPCDQ/LPCDQ status defined. Interestingly, such a time scale is much 
longer than the typical month/year-like intervals observed between the 
nuclear ejections, as mentioned above.  

Detailed characterization of the rapid optical variability has assumed added
relevance in the present Fermi-LAT era (Atwood et al. 2009).
Recent TeV monitoring  has revealed ultra-fast variability on
 minute-like time scales for a few blazars
(Aharonian et al.\ 2007; Albert et al.\ 2007; Acciari et al.\ 2009; Aleksi{\'c} et al.\ 2011).
A scenario proposed
to explain such $\gamma-$ray flaring invokes disturbance caused
in the jet flow by the passage of red giant stars through the inner jet which is
normally opaque to radio emission (Barkov et al.\ 2012). In this mechanism,
continued impact of the jet flow would blow out the extended atmospheres
of such intruding stars, forming magnetized condensations
accelerated to high bulk Lorentz factors. The concomitant shocks at
these condensations
would lead to particle accelaration, accounting for the ultra-rapid TeV flux variations.
Interestingly, this same process would also excite turbulence
in the jet plasma (the process invoked above for the origin of INOV),
powered by the red giants and their wakes crossing the jet.
With a typical stellar velocity of $\la 10^3$ km s$^{-1}$ the expected crossing
time of the inner jet by the star is $\ga 10^2 - 10^3$ yr and so the
postulated enhanced turbulence level in the affected jet sedgement can be
a long lasting feature, consistent with the persistence of enhanced INOV in
blazars underscored in the present work. However, within this basic scenario
it remains to be clearly understood why, unlike the $\gamma$-ray flaring, INOV has been
so rarely detected on sub-hour time scales (e.g., Gopal-Krishna et al.\ 2011).

A potentially very useful tool for constraining INOV mechanism in different
AGN classes is the observation of intra-night variability of polarized light, 
though very few systematic studies have been reported. A preliminary 
investigation by Andruchow, Romero \& Cellone (2005) indicated that at least 
for BL Lac objects, the occurence of optical polarization variability on 
sub-hour times scales is not so rare, unlike the case for optical continuum 
variability (e.g., Gopal-Krishna et al.\ 2011). A more extensive study of 
polarization INOV (`PINOV') has been reported by Villforth et al.\ (2009), 
who monitored an AGN sample consisting of 12 RQQs, 8 BL Lacs and 8 FSRQs, 
albeit for only a single session lasting about 4 hours per AGN. They 
concluded that for sources having  $P_{op} \geq 5\%$, PINOV is ubiquitous but 
it is less frequent among BL Lacs and FSRQs showing lower $P_{op}$. 
Based on this, they have associated PINOV with instabilities in the jet or 
changing physical conditions in the jet plasma.

To sum up, the main conclusion emerging from the present work is that 
compared to HPCDQs the INOV exhibited by LPCDQs is distinctly milder and 
large-amplitude INOV with $\psi > 3$ per cent is very rarely seen for them. 
Given that strong beaming of the nuclear jets is already occuring in both 
HPCDQs and LPCDQs, it would appear from the present work that the effective 
`switch' for strong intranight optical variability is the presence of 
optical polarization, even if its measurement preceded the INOV
observations by several years. To effectively probe this point and the 
connection between INOV and TeV flaring on hour-like or shorter time 
scales, it is important to carry out more sensitive intranight optical 
monitoring of flat-spectrum quasars (both HPCDQs and LPCDQs), preferrably 
in the polarimetric mode and in coordination with their monitoring at TeV 
energies. \\ \\

\begin{acknowledgements} 
The authors thank Dr. Vijay Mohan for help during the observations at IGO and 
Dr. C.S. Stalin for making available his quasar optical monitoring data 
in digital 
form. The authors wish to acknowledge the support received from the staff of the 
IAO and CREST of IIA and IGO. 
We are very thankful to an anonymous referee for 
carefully reviewing the manuscript and for making several constructive suggestions. 
This research has made use of 
NASA/IPAC Extragalactic Database (NED), which is operated by the Jet Propulsion 
Laboratory, California Institute of Technology, under contract with National 
Aeronautics and Space Administration.  
\end{acknowledgements}

\newpage 
\begin{landscape}
\begin{table} 
\centering 
\caption{The LPCDQ and HPCDQ samples studied in the present work$^\$$. \label{tab_cdq_parameter}}
%\tiny 
\begin{tabular}{cccccccrccccc}\\
\hline 
IAU name & Other name   & R.A.(J2000) & Dec(J2000)&{\it B}& $M_{B}$& $z$ &$P_{op}$&$\alpha_{r}$ &  $\alpha_{r,old}$      &   $P_{ext}^{5 GHz}$ &log$f_c$& Ref. \\ 
         &        & (h m s)    &($ ^{\circ}$ $ ^{\prime}$ $ ^{\prime\prime }$) & (mag)  &  (mag)& & ( per cent)   &      &  & (W/Hz) &      &   \\
  (1)    &  (2)  & (3)   & (4)        & (5)   &  (6)   &(7)& (8)        &    (9) &              (10)       &(11)  & (12)           & (13)\\ 
\hline   
\multicolumn{13}{l}{Low-polarization core dominated quasars (LPCDQs)}\\ \hline
J0005$+$0524$^*$ &  UM 18         &00 05 20.1&$+$05 24 11&16.56&$-$26.47 &1.900&1.60$^a$&$-$0.04      &   0.67    &$3.5\times10^{27}$ &0.18$^h$ &(1) \\
J0235$-$0402$^*$ &  PKS 0232$-$02 &02 35 07.2&$-$04 02 05&16.61&$-$26.14&1.458&0.91$^b$&$-$0.49$^\P$  &   0.24    &$<7.6\times10^{28}$ &        &(2) \\
J0456$+$0400$^*$ &PKS 0454$+$039  &04 56 47.1&$+$04 00 53&16.76&$-$25.76 &1.359&0.32$^b$&$+$0.03$^\S$ &   0.11    &$5.6\times10^{27}$ &0.44$^i$ &(2) \\
J0741$+$3111$^*$ & OI 363         &07 41 10.7&$+$31 11 59&17.10&$-$24.34 &0.630&0.44$^c$&$+$0.14$^\S$ &   0.23    &$1.8\times10^{24}$ &3.48$^j$ &(3) \\
J0842$+$1835$^*$& DW 0839$+$18    &08 42 05.1&$+$18 35 42&16.63&$-$25.95 &1.272&1.74$^c$&$-$0.52$^\S$ &   0.17    &$7.6\times10^{27}$ &0.37$^k$ &(3) \\
J0958$+$3224    &    3C 232       &09 58 20.9&$+$32 24 02&15.88&$-$25.40 &0.530&0.53$^c$&$-$0.09$^\S$ &$-$0.27    &$3.2\times10^{26}$ &0.69$^k$ &(4) \\
J1131$+$3114    &  B2 1128$+$31   &11 31 09.4&$+$31 14 07&16.80&$-$23.33 &0.290&0.95$^b$&$-$0.41      &$-$0.21    &$3.6\times10^{25}$ &0.02$^k$ &(4) \\
J1228$+$3128    & B2 1225$+$31    &12 28 24.8&$+$31 28 38&16.15&$-$27.10 &2.219&0.16$^c$&$+$0.01      &   0.0     &$1.5\times10^{27}$ &1.39$^k$ &(4) \\
J1229$+$0203$^*$&   3C 273        &12 29 06.7&$+$02 03 08&13.05&$-$25.88 &0.158&0.50$^e$&$-$0.19$^\S$ &   0.07    &$2.0\times10^{26}$ &1.21$^l$ &(3) \\
J1357$+$1919$^*$& PKS 1354$+$19   &13 57 04.5&$+$19 19 06&16.33&$-$25.27 &0.729&0.43$^c$&$-$0.28$^\S$ &$-$0.23    &$1.9\times10^{27}$ &0.25$^k$ &(3) \\
J2203$+$3145$^*$&  B2 2201$+$31A  &22 03 14.9&$+$31 45 38&15.85&$-$23.90 &0.298&0.72$^c$&$+$0.18$^\S$ &   0.26    &$8.9\times10^{25}$ &1.13$^l$ &(3) \\
J2346$+$0930$^*$& PKS 2344$+$09   &23 46 37.0&$+$09 30 45&16.23&$-$25.21 &0.673&0.90$^b$&$-$0.12$^\S$ &$-$0.08    &$9.1\times10^{26}$ &0.70$^l$ &(2) \\\hline 
\multicolumn{13}{l}{High-polarization core dominated quasars (HPCDQs)}\\ \hline
J0238$+$1637$^*$&AO 0235$+$164    &02 38 38.9&$+$16 37 00&16.46 &$-$25.47 &0.940&43.9$^d$&$+$0.70$^\S$&   0.53    &$2.0\times10^{27}$ &0.83$^l$ &(3) \\
J0423$-$0120$^*$& PKS 0420$-$014  &04 23 15.8&$-$01 20 33&17.50 &$-$24.17 &0.915&20.0$^d$&$+$0.18$^\S$&$-$0.50    &$4.6\times10^{28}$ &0.26$^l$ &(3) \\
J0739$+$0137$^*$&PKS 0736$+$01    &07 39 18.0&$+$01 37 04&16.90 &$-$21.96 &0.191& 5.6$^b$&$-$0.10$^\S$&$-$0.10    &$4.4\times10^{25}$ &0.86$^l$ &(3) \\
J1058$+$0133$^*$& PKS 1055$+$01   &10 58 29.6&$+$01 33 58&18.74 &$-$23.34 &0.888& 5.0$^e$&$+$0.06$^\S$&   0.12    &$1.0\times10^{28}$ &0.57$^l$ &(3) \\
J1159$+$2914$^*$& 4C 29.45        &11 59 31.9&$+$29 14 45.0 &14.41&$-$27.00 &0.729&28.0$^d$&$-$0.34$^\S$& $-$0.42 & $1.2\times10^{27}$ &0.91$^l$ &(3) \\
J1218$-$0119    & PKS 1216$-$010  &12 18 35.0&$-$01 19 54&16.17 &$-$25.14 &0.554& 6.9$^f$&$+$0.01$^\S$&   0.62    &$1.6\times10^{26}$ &0.24$^m$ &(4) \\
J1256$-$0547$^*$& 3C 279          &12 56 11.1&$-$05 47 21&18.01 &$-$23.24 &0.538&44.0$^d$&$+$0.47$^\S$&   0.40    &$1.6\times10^{28}$ &0.42$^l$ &(3) \\
J1310$+$3220    & B2 1308$+$32    &13 10 28.7&$+$32 20 44&15.61 &$-$26.69 &0.997&28.0$^d$&$-$0.09$^\S$&   0.02    &$1.9\times10^{28}$ &0.33$^l$ &(3) \\
J1512$-$0906$^*$&  PKS 1510$-$08  &15 12 50.5&$-$09 06 00&16.74 &$-$23.49 &0.360& 7.8$^d$&$-$0.10$^\S$&   0.78    &$8.5\times10^{25}$ &1.31$^l$ &(5) \\
\hline 
\end{tabular}

$^\$$ Unless otherwise mentioned the observed data are taken from V\'eron-Cetty \& V\'eron (2006). \\
Columns: (1) source name (an asterisk indicates that the CDQ was monitored by us); (2) most popular name as given in
V\'eron-Cetty \& V\'eron (2006); (3) right ascension; (4) declination;
(5) apparent B-magnitude; (6) absolute B-magnitude; (7) redshift;
(8) optical polarization; (9) radio spectral index; 
(10) radio spectral index from V\'eron-Cetty \& V\'eron (1996) (see Sect. \ref{cdq_sample} for explanation);
(11) extended emission radio luminosity at 5 GHz;
(12) radio core dominance fraction, or, core dominance parameter $f_c$ (see text);
(13) Reference for the source selection (see below) .\\
Footnotes: 
Column 8: reference for $P_{op}$: (a) Koratkar et al. \ (1998); (b) Stockman, Moore \& Angel (1984);
(c) Wills et al.\ (1992); (d) Fan et al.\ (1997); (e) Impey \& Tapia (1990);
(f) Visvanathan \& Wills (1998).
Column 9: {$^\S$} radio spectral index derived using the flux measurements from Kovalev et al. (1999) while
the rest are based on the 6 cm and 20 cm fluxes given in V\'eron-Cetty \& V\'eron (2006). {$^\P$}Drinkwater et al.\ (1997).
Column 12: reference for the VLBI fluxes used for estimating $f_c$: (h) Orienti et al.\ (2006);
(i) Briggs (1983); (j) Stanghellini et al.\ (1997); (k)Helmboldt et al.\ (2007);
(l) Lister \& Homan (2005); (m) Wehrle, Morabito \& Preston (1984). For J0235$-$0402 see Sect. 2.1.
Column 13: reference for the source selection (Sect.\ 2) : (1) Koratkar et al.\ (1998); (2) Stockman et al.\ (1984);
(3)Wills et al.\ (1992); (4) Sagar et al.\ (2004); (5) Romero et al.\ (1999).
\end{table}
\end{landscape}

\begin{deluxetable}{ccccccc} 
\tablecolumns{7} 
\tabletypesize{\scriptsize} 
\tablecaption{Positions and magnitudes of the CDQs and the comparison stars.$^*$ \label{tab_cdq_comp}} 
\tablewidth{0pt} 
\tablehead{ 
\colhead{IAU Name} &  Type    &\colhead{R.A.(J2000)} & \colhead{Dec.(J2000)} & \colhead{\it B} & \colhead{\it R} & \colhead{\it B-R} \\  
                   &          &  (h m s)             &($^\circ$ $^\prime$ $^{\prime\prime}$)   &        (mag)    &    (mag)  & (mag)        \\  
\colhead{(1)} & \colhead{(2)} & \colhead{(3)} & \colhead{(4)} & \colhead{(5)} & \colhead{(6)} & \colhead{(7)}  
} 
\startdata 
 
 J0005$+$0524&  LPCDQ     & 00 05 20.21 & $+$05 24 10.9 & 16.51 & 16.26 &0.25\\
          S1 &            & 00 05 32.44 & $+$05 21 07.2 & 17.89 & 16.37 &1.52\\
          S2 &            & 00 04 54.88 & $+$05 28 09.7 & 17.53 & 16.12 &1.41\\
          S3 &            & 00 05 02.42 & $+$05 24 19.6 & 17.27 & 16.34 &0.93\\
          S4 &            & 00 05 27.44 & $+$05 24 45.9 & 17.07 & 16.19 &0.88\\
 J0235$-$0204& LPCDQ      & 02 35 07.34 & $-$04 02 05.2 & 17.13 & 15.94 &1.19\\
          S1 &            & 02 35 16.05 & $-$03 59 52.1 & 15.98 & 15.60 &0.38\\
          S2 &            & 02 35 21.59 & $-$04 08 11.1 & 16.31 & 15.32 &0.99\\
          S3 &            & 02 35 00.40 & $-$04 07 25.6 & 17.12 & 15.43 &1.69\\
          S4 &            & 02 35 07.76 & $-$04 00 23.9 & 17.92 & 16.23 &1.69\\
 J0456$+$0400&  LPCDQ     & 04 56 47.16 & $+$04 00 53.0 & 16.69 & 16.26 &0.43\\
          S1 &            & 04 56 28.46 & $+$04 00 55.5 & 15.62 & 15.04 &0.58\\
          S2 &            & 04 56 28.75 & $+$04 01 30.0 & 15.96 & 15.37 &0.59\\
          S3 &            & 04 56 50.81 & $+$04 00 31.1 & 17.00 & 15.88 &1.12\\
 J0741$+$3112&  LPCDQ     & 07 41 10.69 & $+$31 12 00.4 & 16.65 & 16.29 &0.36\\
          S1 &            & 07 41 24.15 & $+$31 09 44.8 & 16.06 & 14.82 &1.24\\
          S2 &            & 07 41 20.71 & $+$31 08 49.8 & 16.19 & 15.09 &1.10\\
          S3 &            & 07 41 00.69 & $+$31 16 44.4 & 16.67 & 15.57 &1.10\\
          S4 &            & 07 41 07.97 & $+$31 11 48.6 & 16.65 & 15.60 &1.05\\
 J0842$+$1835&  LPCDQ     & 08 42 05.09 & $+$18 35 41.1 & 17.59 & 16.56 &1.03\\
          S1 &            & 08 42 21.26 & $+$18 35 26.8 & 18.25 & 16.13 &2.12\\
          S2 &            & 08 42 28.18 & $+$18 37 28.4 & 17.61 & 15.58 &2.03\\
          S3 &            & 08 42 26.06 & $+$18 36 27.1 & 16.54 & 15.36 &1.18\\
 J1229$+$0203&  LPCDQ     & 12 29 06.70 & $+$02 03 08.5 & 13.73 & 14.11 &-0.38\\
          S1 &            & 12 29 03.20 & $+$02 03 18.8 & 14.12 & 13.42 &0.70 \\
          S2 &            & 12 28 50.92 & $+$02 06 31.4 & 13.22 & 12.32 &0.90\\
          S3 &            & 12 29 08.39 & $+$02 00 18.7 & 13.39 & 12.10 &1.29\\
 J1357$+$1919&  LPCDQ     & 13 57 04.43 & $+$19 19 07.5 & 16.59 & 16.29 &0.30\\
          S1 &            & 13 57 04.60 & $+$19 20 24.2 & 16.91 & 15.61 &1.30\\
          S2 &            & 13 57 07.00 & $+$19 22 30.0 & 17.92 & 15.85 &2.07\\
          S3 &            & 13 57 19.36 & $+$19 17 57.7 & 17.43 & 15.66 &1.77\\
          S4 &            & 13 56 52.13 & $+$19 20 51.8 & 16.49 & 14.91 &1.58\\
          S5 &            & 13 56 52.79 & $+$19 14 59.2 & 16.49 & 15.22 &1.27\\
 J2203$+$3145&  LPCDQ     & 22 03 14.97 & $+$31 45 38.4 & 15.39 & 14.33 &1.06\\
          S1 &            & 22 02 58.00 & $+$31 48 43.3 & 16.00 & 15.05 &0.95\\
          S2 &            & 22 03 27.10 & $+$31 41 47.4 & 15.86 & 14.64 &1.22\\
          S3 &            & 22 02 52.30 & $+$31 46 51.2 & 15.75 & 15.03 &0.72\\
          S4 &            & 22 02 56.91 & $+$31 44 50.3 & 16.36 & 15.60 &0.76\\
 J2346$+$0930&  LPCDQ     & 23 46 36.82 & $+$09 30 45.8 & 16.34 & 15.99 &0.35\\
          S1 &            & 23 46 47.90 & $+$09 25 59.6 & 18.66 & 16.57 &2.09\\
          S2 &            & 23 46 53.42 & $+$09 26 10.6 & 17.25 & 14.94 &2.31\\
          S3 &            & 23 46 53.56 & $+$09 29 20.7 & 17.26 & 15.99 &1.27\\
          S4 &            & 23 46 22.91 & $+$09 29 35.6 & 16.00 & 14.87 &1.13\\
             &            &             &               &       &       &    \\
 J0238$+$1637&  HPCDQ     & 02 38 38.92 & $+$16 36 59.2 & 18.65 & 15.92 &2.73\\
          S1 &            & 02 38 56.00 & $+$16 37 43.0 & 17.43 & 16.60 &0.83\\
          S2 &            & 02 38 38.52 & $+$16 40 05.3 & 18.37 & 16.61 &1.76\\
          S3 &            & 02 38 22.25 & $+$16 39 41.8 & 17.37 & 16.22 &1.15\\
 J0423$-$0120& HPCDQ      & 04 23 15.79 & $-$01 20 33.1 & 15.62 & 16.28 &-0.66\\
          S1 &            & 04 22 57.47 & $-$01 18 02.0 & 15.87 & 15.27 &0.60\\
          S2 &            & 04 23 08.03 & $-$01 18 58.2 & 16.09 & 15.65 &0.44\\
          S3 &            & 04 23 11.50 & $-$01 18 23.6 & 16.96 & 15.86 &1.10\\
          S3 &            & 04 23 15.17 & $-$01 22 39.4 & 16.53 & 15.74 &0.79\\
 J0739$+$0137& HPCDQ      & 07 39 18.03 & $+$01 37 04.6 & 16.27 & 16.19 &0.08\\
          S1 &            & 07 39 13.09 & $+$01 32 28.7 & 15.93 & 15.50 &0.43\\
          S2 &            & 07 39 10.65 & $+$01 36 43.6 & 15.94 & 16.20 &-0.26\\
          S3 &            & 07 39 14.30 & $+$01 33 18.4 & 15.95 & 15.77 &0.18\\
 J1058$+$0133& HPCDQ      & 10 58 29.60 & $+$01 33 58.9 & 18.00 & 16.68 &1.32\\
          S1 &            & 10 58 27.43 & $+$01 34 33.2 & 16.65 & 15.22 &1.43\\
          S2 &            & 10 58 33.73 & $+$01 29 52.9 & 16.90 & 15.23 &1.67\\
          S3 &            & 10 58 11.16 & $+$01 28 20.6 & 15.83 & 14.31 &1.52\\
 J1159$+$2914& HPCDQ      & 11 59 31.8  & $+$29 14 43.9 & 17.45 & 17.39 &0.06\\
          S1 &            & 11 59 39.11 & $+$29 17 54.9 & 16.26 & 17.43 &-1.17\\ 
          S2 &            & 11 59 53.61 & $+$29 15 49.4 & 16.96	& 16.28 &0.68\\ 
          S3 &            & 11 59 27.09 & $+$29 16 31.1 & 18.15 & 16.88 &1.27\\ 
 J1256$-$0547& HPCDQ      & 12 56 11.19 & $-$05 47 21.5 & 17.39 & 15.87 &1.52\\
          S1 &            & 12 56 26.61 & $-$05 45 22.8 & 15.22 & 14.75 &0.47\\
          S2 &            & 12 55 58.00 & $-$05 44 18.9 & 16.19 & 15.30 &0.89\\
          S3 &            & 12 56 14.48 & $-$05 46 47.8 & 16.39 & 15.43 &0.96\\
 J1512$-$0906& HPCDQ      & 15 12 50.54 & $-$09 05 59.7 & 16.72 & 15.93 &0.79\\
          S1 &            & 15 12 41.21 & $-$09 06 34.5 & 16.42 & 14.54 &1.88\\
          S2 &            & 15 12 59.18 & $-$09 10 31.4 & 16.09 & 15.07 &1.02\\
          S3 &            & 15 13 08.88 & $-$09 02 33.8 & 16.64 & 15.14 &1.50\\
 
\enddata 

$^*$ Taken from Monet et al. (2003)

\end{deluxetable} 
\clearpage
\newpage

\begin{deluxetable}{cccccccccccccc} 
\tablecolumns{13}
\tabletypesize{\tiny}
\tablecaption{Summary of observations and  derived INOV parameters for the LPCDQ sample \label{tab_inov_lpcdq}} 
\tablewidth{0pt}
\tablehead{
Source  & Date    & Tel.{$^\P$}  & Dur. & $N_p$ & $\sigma $ & $\psi$ & $\Delta m_{CS1}, \Delta m_{CS2}    $  &  $F_{CS1}, F_{CS2}$&  Status$^\dag$ & $F_{CS1-CS2}$   &  Status$^\dag$  & Ref.$^\pounds$  \\
        &  dd.mm.yy& used           &   (hrs)  &     & (\%)      &  (\%)  &         &     &  $F_{CS1}, F_{CS2}$         &     &   (CS1-CS2)      &                 \\
        (1) & (2) & (3) & (4) & (5) & (6) & (7) & (8) & (9) & (10) & (11)   & (12) & (13)     \\ }
\startdata

J0005$+$0524  & 23.10.06 &  ST  &6.0     & 14      &  0.11   &   0.95  & -0.06, -0.19  & 2.76 , 3.72  & PV,PV  &  0.21    &   N     &   (a)  \\
              & 18.11.06 &  ST  &3.9     & 09      &  0.06   &   0.53  &  0.09,  0.30  & 1.16 , 2.63  &  N,N   &  0.11    &   N     &   (a)  \\
              & 14.09.07 &  ST  &4.3     & 10      &  0.15   &   0.72  & -0.10,  0.26  & 1.00 , 1.46  &  N,N   &  0.60    &   N     &   (a)  \\
              & 16.09.07 &  ST  &5.2     & 11      &  0.20   &   1.34  &  0.03, -0.23  & 6.90 , 2.25  &  V,N   &  0.93    &   N     &   (a)  \\
              &          &      &        &         &         &         &               &              &        &          &         &        \\
J0235$-$0402  & 21.10.04 &  ST  &6.3     & 13      &  0.17   &   1.43  & 0.70, 0.83    & 7.20 , 3.76  &  V,PV  &  1.85    &   N     &   (a)  \\
              & 22.10.04 &  ST  &6.7     & 15      &  0.11   &   0.89  & 0.12, 0.83    & 1.03 , 1.51  &  N,N   &  0.33    &   N     &   (a)  \\
              & 04.11.04 &  ST  &5.7     & 23      &  0.13   &   0.88  & 0.59, 0.84    & 1.69 , 2.63  &  N,PV  &  0.74    &   N     &   (a)  \\
              & 05.11.04 &  ST  &6.8     & 27      &  0.13   &   0.45  & 0.71, 0.83    & 0.37 , 0.60  &  N,N   &  1.09    &   N     &   (a)  \\
              &          &      &        &         &         &         &               &              &        &          &         &        \\
J0456$+$0400  & 23.11.08 &  ST  &6.0     & 22      &  0.17   &   1.67  & 0.41, 1.39    & 2.36 , 3.83  & PV, V  &  0.93    &   N     &   (a)  \\
              & 29.11.08 &  ST  &5.0     & 18      &  0.12   &   0.85  & 0.99, 1.39    & 1.05 , 0.91  & N,  N  &  0.37    &   N     &   (a)  \\
              & 03.12.08 &  ST  &4.9     & 20      &  0.20   &   1.17  & 0.46, 0.99    & 1.22 , 0.92  & N,  N  &  0.67    &   N     &   (a)  \\
              &          &      &        &         &         &         &               &              &        &          &         &        \\
J0741$+$3112  & 20.01.06 &  ST  &7.0     & 29      &  0.16   &   0.72  & 0.46, 1.07    & 0.92 , 2.25  & N, PV  &  1.08    &   N     &   (a)  \\
              & 21.01.06 &  ST  &3.6     & 16      &  0.15   &   4.88  & 0.20, 0.96    & 35.35, 61.39 & V, V   &  0.50    &   N     &   (a)  \\
              & 18.12.06 &  ST  &6.8     & 28      &  0.10   &   0.95  & 1.05, 1.19    & 1.29 , 1.17  & N, N   &  0.69    &   N     &   (a)  \\
              & 22.12.06 &  ST  &7.3     & 30      &  0.11   &   1.33  & 1.04, 1.18    & 3.30 , 3.27  & V, V   &  0.79    &   N     &   (a)  \\
              &          &      &        &         &         &         &               &              &        &          &         &        \\
J0842$+$1835  & 04.02.06 &  ST  &7.1     & 26      &  0.14   &   3.44  & 1.33, 1.61    & 11.09, 10.2  & V, V   &  1.39    &   N     &   (a)  \\
              & 16.12.06 &  ST  &5.0     & 12      &  0.18   &   1.68  & 0.81, 1.36    & 2.29 , 1.50  & N, N   &  0.90    &   N     &   (a)  \\
              & 21.12.06 &  ST  &6.5     & 28      &  0.12   &   1.46  & 0.79, 1.34    & 1.83 , 2.59  & N, V   &  0.36    &   N     &   (a)  \\
              &          &      &        &         &         &         &               &              &        &          &         &        \\
J0958$+$3224  & 19.02.99 &  ST  &6.5     & 34      &  0.22   &   1.21  &  -0.48, 1.24  & 0.68 , 1.37  & N, N   &  0.34    &   N     &   (b)  \\
              & 03.03.00 &  ST  &6.3     & 35      &  0.32   &   0.83  &  -0.54, 0.77  & 0.63 , 1.82  & N, PV  &  1.49    &   N     &   (b)  \\
              & 05.03.00 &  ST  &6.9     & 32      &  0.16   &   0.66  &  -0.54, 0.77  & 0.56 , 1.24  & N, N   &  0.34    &   N     &   (b)  \\
              &          &      &        &         &         &         &               &              &        &          &         &        \\
J1131$+$3114  & 18.01.01 &  ST  &5.7     & 29      &  0.21   &   0.72  &-0.10, 0.13    & 0.77 , 0.82  & N, N   &  0.72    &   N     &   (b)  \\
              & 09.03.02 &  ST  &8.2     & 25      &  0.24   &   1.22  & 0.20, -0.23   & 1.79 , 1.01  & N, N   &  0.91    &   N     &   (b)  \\
              & 10.03.02 &  ST  &8.3     & 26      &  0.19   &   0.45  & 0.00, 0.20    & 0.33 , 1.12  & N, N   &  0.93    &   N     &   (b)  \\
              &          &      &        &         &         &         &               &              &        &          &         &        \\
J1228$+$3128  & 07.03.99 &  ST  &6.6     & 47      &  0.43   &   1.82  & -0.19,  1.10  & 1.67 , 2.71  & PV, V  &  2.27    &   V     &   (b)  \\
              & 07.04.00 &  ST  &6.0     & 25      &  0.57   &   1.54  & -0.20, -1.56  & 1.70 , 0.76  & N, N   &  1.24    &   N     &   (b)  \\
              & 20.04.01 &  ST  &7.4     & 32      &  0.40   &   1.48  & -0.21, -1.57  & 2.21 , 1.52  & PV, N  &  0.56    &   N     &   (b)  \\
              &          &      &        &         &         &         &               &              &        &          &         &        \\
J1229$+$0203  & 07.03.11 &  ST  &4.9     & 32      &  0.14   &   0.91  &  0.26, 0.34   & 3.92 , 2.25  & V, PV  &  1.27    &   N     &   (a)  \\
              & 10.03.11 &  ST  &6.3     & 47      &  0.18   &   0.67  &  0.24, 0.28   & 1.29 , 1.42  & N, N   &  2.20    &   V     &   (a)  \\
              & 09.04.11 &  IGO &5.6     & 49      &  0.13   &   0.68  &  0.27, 0.28   & 2.81 , 2.31  & V, V   &  1.68    &   PV    &   (a)  \\
              &          &      &        &         &         &         &               &              &        &          &         &        \\
J1357$+$1919  & 27.02.06 &  ST  &4.2     & 10      &  0.11   &   1.25  & 0.34, 0.35    & 13.38, 17.79 & V, V   &  0.78    &   N     &   (a)  \\
              & 05.03.06 &  ST  &4.0     & 09      &  0.12   &   0.53  & 0.30, 0.32    & 1.71 , 4.46  & N, PV  &  1.58    &   N     &   (a)  \\
              & 26.03.06 &  ST  &5.8     & 10      &  0.20   &   0.53  & 0.29,  0.34   & 1.06 , 0.71  & N, N   &  2.78    &   N     &   (a)  \\
              & 28.03.06 &  ST  &5.2     & 18      &  0.17   &   3.58  & 0.29,  0.35   & 33.51, 41.42 & V, V   &  1.22    &   N     &   (a)  \\
              & 29.03.06 &  ST  &5.3     & 19      &  0.21   &   0.41  & 0.29,  0.32   & 0.47 , 0.63  & N, N   &  2.36    &   PV    &   (a)  \\
              & 06.04.06 &  ST  &6.8     & 24      &  0.20   &   1.12  & 0.33,  0.35   & 2.43 , 1.18  & PV, N  &  1.03    &   N     &   (a)  \\
              & 22.04.06 &  ST  &4.1     & 15      &  0.20   &   0.60  & 0.31,  0.35   & 0.88 , 1.62  & N, N   &  0.90    &   N     &   (a)  \\
              & 23.04.06 &  ST  &4.4     & 14      &  0.28   &   2.16  & 0.79, 1.58    & 4.80 , 4.90  & V, V   &  1.39    &   N     &   (a)  \\
              &          &      &        &         &         &         &               &              &        &          &         &        \\
J2203$+$3145  & 08.11.05 &  HCT &3.6     & 15      &  0.06   &   0.81  & -0.04, -0.35  & 5.17 , 5.94  & V, V   &  0.23    &   N     &   (a)  \\
              & 14.09.06 &  ST  &5.4     & 24      &  0.20   &   0.90  & -0.15, 0.33   & 3.69 , 5.19  & V, V   &  2.70    &   PV    &   (a)  \\
              & 15.09.07 &  ST  &7.1     & 30      &  0.09   &   0.58  & -0.04, 0.50   & 1.12 , 1.22  & N, N   &  0.35    &   N     &   (a)  \\
              &          &      &        &         &         &         &               &              &        &          &         &        \\
J2346$+$0930  & 20.09.03 & HCT  &5.3     & 37      &  0.16   &   1.71  & -0.67, 0.81   & 7.69 , 19.49 & V, V   &  0.66    &   N     &   (a)  \\
              & 20.10.04 &  ST  &5.1     & 10      &  0.16   &   0.74  &  0.73, 0.86   & 3.31 , 3.69  & PV, PV &  1.73    &   N     &   (a)  \\
              & 16.11.06 &  ST  &4.3     & 10      &  0.10   &   0.32  &  0.23, 0.97   &  0.79, 0.43  & N, N   &  0.55    &   N     &   (a)  \\
\enddata

Columns :- (1) source name; (2) date of observation; (3) telescope used; (4) duration of
monitoring; (5) number of data points in the DLC; (6) rms of the steadiest star-star DLC; 
(7) INOV amplitude ($\psi$); (8) mean magnitude differences: (Q-CS1) and (Q-CS2) for the 
night; (9) $F$-values computed for the Q -CS1 and Q -CS2 DLCs; (10) variability status 
estimated from the $F_{CS1}, F_{CS2}$ values, respectively; (11) $F$-value for the (CS1-CS2)
DLC; (12) variability status for the (CS1-CS2) DLC; 
(13) reference for the INOV data (See text for more information Sect. 4.1).

{$^\P$} ST - Sampurnanad Telescope (ARIES); HCT - Himalayan Chandra Telescope (IIA);
IGO - IUCAA Girawali Observatory. \\
$^\dag$ V = Variable; N = Non-variable; PV = Probable Variable; \\
{$^\pounds$}References for the INOV data: (a) Sagar et al.\ (2004); (b) present work .
\end{deluxetable}

\clearpage

\newpage
\begin{deluxetable}{cccccccccccccccc}
\tablecolumns{13}
\tabletypesize{\tiny}
\tablecaption{Summary of observations and derived INOV parameters for HPCDQ sample \label{tab_inov_hpcdq}}
\tablewidth{0pt}
\tablehead{
Source  & Date    & Tel.{$^\P$}  & Dur. & $N_p$ & $\sigma $ & $\psi$ & $\Delta m_{CS1}, \Delta m_{CS2}    $  &  $F_{CS1}, F_{CS2}$&  Status$^\dag$ & $F_{CS1-CS2}$   &  Status$^\dag$  & Ref.$^\pounds$  \\
        &  dd.mm.yy& used           &   (hrs)  &     & (\%)      &  (\%)  &         (mag, mag)  &    &  $F_{CS1}, F_{CS2}$    &     &   (CS1-CS2)      &                 \\
        (1) & (2) & (3) & (4) & (5) & (6) & (7) & (8) & (9) & (10) & (11)   & (12) & (13)    \\ }
\startdata

J0238$+$1637  & 12.11.99 &  ST   &6.6       & 38      &  0.42   &    12.26 & 0.65, 1.67 & 29.68  ,   34.47 &  V, V        &   1.28    &    N      &(b)  \\
              & 14.11.99 &  ST   &6.2       & 32      &  0.24   &    8.70  & 2.50, 3.50 & 5.27   ,   5.88  &  V, V        &   1.61    &    N      &(b)  \\
              & 18.11.03 & HCT   &7.4       & 39      &  0.30   &    7.31  & 0.50, 0.75 & 36.21  ,   36.63 &  V, V        &   0.99    &    N      &(a)  \\
              &          &       &          &         &         &          &            &                  &              &           &           &     \\
J0423$-$0120  & 19.11.03 & HCT   &6.3       & 36      &  0.18   &    1.68  & -0.10, 0.29& 11.05  ,   14.38 &  V, V        &   1.51    &    N      &(a)  \\
              & 08.12.04 &  ST   &6.0       & 11      &  0.21   &    1.90  & 1.82, 2.23 & 1.84   ,   2.71  &  N, N        &   2.64    &    N       &(a)  \\
              & 25.10.09 &  ST   &4.0       & 18      &  0.34   &    2.74  & 0.93, 1.11 & 3.87   ,   5.15  &  V, V        &   1.25    &    N       &(a)  \\
              &          &       &          &         &         &          &            &                  &              &           &           &     \\
J0739$+$0137  & 05.12.05 & HCT   &5.3       & 10      &  0.21   &    3.75  & 1.34, 1.80 & 9.72   ,   10.32 &  V, V        &   0.99    &    N       &(a)  \\
              & 06.12.05 & HCT   &6.0       &  9      &  0.44   &    2.86  & 1.19, 1.83 & 9.90   ,   11.03 &  V, V        &   4.47    &    PV      &(a)  \\
              & 09.12.05 & HCT   &5.5       & 14      &  0.29   &    0.88  & 1.22, 1.40 & 0.28   ,   0.41  &  N, N        &   1.97    &    N       &(a)  \\
              &          &       &          &         &         &          &            &                  &              &           &           &     \\
J1058$+$0133  & 25.03.07 &  ST   &5.8       & 11      &  0.08   &    2.08  & 0.68, 0.86 & 9.74   ,   8.64  &  V, V        &   0.45    &    N       &(a)  \\
              & 16.04.07 &  ST   &3.8       & 15      &  0.17   &    0.52  & 0.72, 1.62 & 0.49   ,   1.12  &  N, N        &   0.85    &    N       &(a)  \\
              & 23.04.07 &  ST   &4.4       & 10      &  0.17   &    1.59  & 0.47, 0.63 & 6.11   ,   5.56  &  V, V        &   0.67    &    N       &(a)  \\
              &          &       &          &         &         &          &            &                  &              &           &           &     \\
J1159$+$2914  & 31.03.12 &  ST   &5.1       & 16      &  0.47   &   5.96   & 0.57, 0.71 & 4.20   ,   3.51  &  V, V        &   0.45    &    N       &(a)  \\
              & 01.04.12 &  ST   &7.5       & 23      &  0.40   &   9.73   & 0.56, 0.69 & 11.65  ,   11.45 &  V, V        &   0.53    &    N       &(a)  \\
              & 02.04.12 &  ST   &6.6       & 19      &  2.11   &   16.35  & -0.02, 0.11& 17.95  ,   22.76 &  V, V        &   2.56    &    PV      &(a)  \\
              &          &       &          &         &         &          &            &                  &              &           &           &     \\
J1218$-$0119  & 11.03.02 &  ST   &8.0       & 20      &  0.18   &    4.58  & 1.36, 1.44 & 5.47   ,   5.91  &  V, V        &   0.70    &    N       &(b)  \\
              & 13.03.02 &  ST   &8.5       & 22      &  0.29   &    3.10  & 1.34, 1.42 & 3.76   ,   4.83  &  V, V        &   2.13    &    PV      &(b)  \\
              & 15.03.02 &  ST   &3.9       &  9      &  0.13   &    2.45  & 1.45, 1.53 & 6.35   ,   7.84  &  V, V        &   0.56    &    N       &(b)  \\
              & 16.03.02 &  ST   &8.2       & 20      &  0.22   &    13.02 & 1.31, 1.39 & 154.12 ,   166.24&  V, V        &   2.26    &    V        &(b)  \\
              &          &       &          &         &         &          &            &                  &              &           &           &     \\
J1256$-$0547  & 26.01.06 &  ST   &4.2       & 19      &  0.17   &    2.49  & -0.07,-0.44& 28.63  ,   33.73 &  V, V        &   1.81    &    N        &(a)  \\
              & 28.02.06 &  ST   &6.1       & 40      &  0.15   &    10.26 & -0.32,-0.92& 619.23 ,   539.22&  V, V        &   1.33    &    N       &(a)  \\
              & 20.04.09 &  ST   &4.9       & 20      &  0.23   &    22.05 & 1.14, 2.10 & 172.63 ,   183.19&  V, V        &   1.27    &    N       &(a)  \\
              &          &       &          &         &         &          &            &                  &              &           &           &     \\
J1310$+$3220  & 26.04.00 &  ST   &5.6       & 16      &  0.34   &   1.43   &0.97,  1.01 & 0.16   ,   0.19  &  N, N        &   0.19    &    N       &(b)  \\
              & 17.03.02 &  ST   &7.7       & 19      &  0.35   &   3.30   &0.12, -0.92 & 13.23  ,   3.46  &  V, V        &   0.39    &    N       &(b)  \\
              & 24.04.02 &  ST   &5.8       & 12      &  0.14   &   0.33   &-0.48, 0.55 & 0.14   ,   0.54  &  N, N        &   0.12    &    N       &(b)  \\
              & 02.05.02 &  ST   &5.1       & 13      &  0.60   &   1.14   & 0.49, 0.52 & 0.43   ,   0.21  &  N, N        &   0.26    &    N       &(b)  \\
              &          &       &          &         &         &          &            &                  &              &           &           &     \\
J1512$-$0906  & 14.06.05 &  ST   &4.0       &  9      &  0.17   &   1.55   & 1.60, 2.17 & 2.91   ,   2.73  &  N, N        &   1.94    &    N       &(a)  \\
              & 01.05.09 &  ST   &5.6       & 22      &  0.26   &   5.33   & 0.43, 0.46 & 12.65  ,   9.89  &  V, V        &   0.65    &    N       &(a)  \\
              & 20.05.09 &  ST   &4.8       & 23      &  0.40   &   3.00   & 0.61, 0.63 & 1.41   ,   2.11  &  N, PV       &   0.61    &    N       &(a)  \\
\enddata

Columns :- (1) source name; (2) epoch of observation; (3) telescope used; (4) duration of monitoring;
(5) number of data points in the DLC; (6) rms of the steadiest star-star DLC; (7) INOV amplitude ($\psi$);
(8) mean magnitude difference: Q-CS1 and Q-CS2 for the night;          
(9) $F$-values computed for the Q -CS1 and Q -CS2 DLCs; (10) variability status estimation for $F_{CS1}, F_{CS2}$values, respectively;             
(11) $F$-value for the CS1-CS2 DLC ; (12) variability status for the CS1-CS2 DLC; 
(13) reference for the INOV data (See text for more information Sect. 4.1).

{$^\P$} ST - Sampurnanad Telescope (ARIES); HCT - Himalayan Chandra Telescope (IIA);
IGO - IUCAA Girawali Observatory. \\
$^\dag$ V = Variable; N = Non-variable; PV = Probable Variable\\
{$^\pounds$}References for the INOV data: (a) present work; (b) Sagar et al.\ (2004).

\end{deluxetable}

\clearpage

\newpage
\begin{table*}  
\centering  
\caption{Estimates of DC for our sets of LPCDQs and HPCDQs (using the chosen 2 comparison
stars).} 
\label{dc_stat} 
\begin{tabular}{ccc}\\ 
\hline 
                          & INOV DC   &  INOV DC   \\
                                & (using CS1) &  (using CS2) \\
                                &  (per cent) & (per cent)  \\
\hline                                                                                                         
{\bf LPCDQs}      &             &            \\
for all values of $\psi$ :      & 28(45)$^\dag$& 28(46)$^\dag$  \\
for $\psi > 3$ per cent :       & 7           & 7       \\
{\bf HPCDQs}      &             &            \\
for all values of $\psi$ :      & 68(68)$^\dag$      & 68(72)$^\dag$  \\
for $\psi > 3$ per cent :       & 40          & 40      \\
                                &             &         \\
\hline 
\end{tabular} 

$^\dag$Values inside parentheses are when `PV' cases are also included.

\end{table*}

\begin{table*}  
\centering  
\caption{Results of the two-sample parameter K-S test performed on various 
parameters of our sets of 12 LPCDQs and 9 HPCDQs (Sect. ~\ref{discussion}).} 
\label{ks-test_cdq} 
\begin{tabular}{ccc}\\ 
\hline 
 Parameter     & $d-$statistic & Probability    \\
\hline 
$z  $          & 0.41           &  0.25          \\
$M_B$          & 0.41           &  0.25          \\
$\alpha_r$     & 0.39           &  0.33          \\
$f_{c}   $     & 0.25           &  0.85          \\
$P_{ext }$     & 0.44           &  0.20          \\
$P_{op}  $     & 1.00           &  $1.4 \times 10^{-5} $  \\
\hline 
\end{tabular} 
\end{table*} 
 
\clearpage
\newpage

\begin{figure*} 
\hbox{ 
\hspace*{-1.3cm} 
\includegraphics[height=10.0cm,width=04.4cm]{fig_J0005+0524_23oct06_ST.epsi} 
\hspace*{0.1cm} 
\includegraphics[height=10.0cm,width=04.4cm]{fig_J0005+0524_18nov06_ST.epsi} 
\hspace*{0.1cm} 
\includegraphics[height=10.0cm,width=04.4cm]{fig_J0005+0524_14sep07_ST.epsi} 
\hspace*{0.1cm} 
\includegraphics[height=10.0cm,width=04.4cm]{fig_J0005+0524_16sep07_ST.epsi}  
} 
\vspace*{50pt} 
\hbox{ 
\hspace*{-1.3cm} 
\includegraphics[height=10.0cm,width=04.4cm]{fig_J0235-0402_21oct04_ST.epsi} 
\hspace*{0.1cm} 
\includegraphics[height=10.0cm,width=04.4cm]{fig_J0235-0402_22oct04_ST.epsi} 
\hspace*{0.1cm} 
\includegraphics[height=10.0cm,width=04.4cm]{fig_J0235-0402_04nov04_ST.epsi} 
\hspace*{0.1cm} 
\includegraphics[height=10.0cm,width=04.4cm]{fig_J0235-0402_05nov04_ST.epsi} 
} 
\caption{ 
The intranight optical DLCs of the LPCDQs monitored in the present study. 
For each night, the source name, the telescope used, 
the date, and the duration of monitoring are given at the top. 
The upper 3 panels show the DLCs of the LPCDQ relative to 3 
comparison stars while the attached lower 3 panels show the star-star DLCs, where  
the solid horizontal lines mark the mean for each star-star DLC. 
The bottom panel gives the plots of seeing variation for the night, 
based on 3 stars monitored along with the blazar on the same CCD frame.
} 
\label{fig:1} 
\end{figure*} 

\newpage 
\begin{figure*} 
\hbox{ 
\hspace*{-1.3cm} 
\includegraphics[height=10.0cm,width=04.4cm]{fig_J0456+0400_23nov08_ST.epsi} 
\hspace*{0.1cm} 
\includegraphics[height=10.0cm,width=04.4cm]{fig_J0456+0400_29nov08_ST.epsi} 
\hspace*{0.1cm} 
\includegraphics[height=10.0cm,width=04.4cm]{fig_J0456+0400_03dec08_ST.epsi} 
\hspace*{0.1cm} 
\includegraphics[height=10.0cm,width=04.4cm]{fig_J0741+3112_20jan06_ST.epsi}  
} 
\vspace*{50pt} 
\hbox{ 
\hspace*{-1.3cm} 
\includegraphics[height=10.0cm,width=04.4cm]{fig_J0741+3112_21jan06_ST.epsi} 
\hspace*{0.1cm} 
\includegraphics[height=10.0cm,width=04.4cm]{fig_J0741+3112_18dec06_ST.epsi} 
\hspace*{0.1cm} 
\includegraphics[height=10.0cm,width=04.4cm]{fig_J0741+3112_22dec06_ST.epsi} 
\hspace*{0.1cm} 
\includegraphics[height=10.0cm,width=04.4cm]{fig_J0842+1835_04feb06_ST.epsi} 
} 
\begin{center} 
{{\bf Figure~\ref{fig:1}}. \textit {continued}} 
\end{center} 
\end{figure*}  
  
\newpage 
\begin{figure*} 
\hbox{ 
\hspace*{-1.3cm} 
\includegraphics[height=10.0cm,width=04.4cm]{fig_J0842+1835_16dec06_ST.epsi} 
\hspace*{0.1cm} 
\includegraphics[height=10.0cm,width=04.4cm]{fig_J0842+1835_21dec06_ST.epsi} 
\hspace*{0.1cm} 
\includegraphics[height=10.0cm,width=04.4cm]{fig_J1229+0203_07mar11_ST.epsi} 
\hspace*{0.1cm} 
\includegraphics[height=10.0cm,width=04.4cm]{fig_J1229+0203_10mar11_ST.epsi} 
} 
\vspace*{50pt} 
\hbox{ 
\hspace*{-1.3cm} 
\includegraphics[height=10.0cm,width=04.4cm]{fig_J1229+0203_09apr11_IGO.epsi} 
\hspace*{0.1cm} 
\includegraphics[height=10.0cm,width=04.4cm]{fig_J1357+1919_27feb06_ST.epsi} 
\hspace*{0.1cm} 
\includegraphics[height=10.0cm,width=04.4cm]{fig_J1357+1919_05mar05_ST.epsi} 
\hspace*{0.1cm} 
\includegraphics[height=10.0cm,width=04.4cm]{fig_J1357+1919_26mar06_ST.epsi} 
} 
\begin{center} 
{{\bf Figure~\ref{fig:1}}. \textit {continued}} 
\end{center} 
 
\end{figure*} 
\newpage 
 
\begin{figure*} 
\hbox{ 
\hspace*{-1.3cm} 
\includegraphics[height=10.0cm,width=04.4cm]{fig_J1357+1919_28mar06_ST.epsi} 
\hspace*{0.1cm} 
\includegraphics[height=10.0cm,width=04.4cm]{fig_J1357+1919_29mar06_ST.epsi} 
\hspace*{0.1cm} 
\includegraphics[height=10.0cm,width=04.4cm]{fig_J1357+1919_06apr06_ST.epsi} 
\hspace*{0.1cm} 
\includegraphics[height=10.0cm,width=04.4cm]{fig_J1357+1919_22apr06_ST.epsi} 
} 
\vspace*{50pt} 
\hbox{ 
\hspace*{-1.3cm} 
\includegraphics[height=10.0cm,width=04.4cm]{fig_J1357+1919_23apr06_ST.epsi} 
\hspace*{0.1cm} 
\includegraphics[height=10.0cm,width=04.4cm]{fig_J2203+3145_08nov05_HCT.epsi} 
\hspace*{0.1cm} 
\includegraphics[height=10.0cm,width=04.4cm]{fig_J2203+3145_14sep06_ST.epsi} 
\hspace*{0.1cm} 
\includegraphics[height=10.0cm,width=04.4cm]{fig_J2203+3145_15sep07_ST.epsi} 
} 
\begin{center} 
{{\bf Figure~\ref{fig:1}}. \textit {continued}} 
\end{center} 
\end{figure*} 

\newpage
\begin{figure*}
\hbox{
\hspace*{-1.3cm}
\includegraphics[height=10.0cm,width=04.4cm]{fig_J2346+0930_20sep03_HCT.epsi} 
\hspace*{0.1cm} 
\includegraphics[height=10.0cm,width=04.4cm]{fig_J2346+0930_20oct04_ST.epsi} 
\hspace*{0.1cm} 
\includegraphics[height=10.0cm,width=04.4cm]{fig_J2346+0930_16nov06_ST.epsi} 
\hspace*{0.1cm} 
}
\begin{center}
{{\bf Figure~\ref{fig:1}}. \textit {continued}}
\end{center}
\end{figure*}

\newpage 
\begin{figure*} 
\hbox{ 
\hspace*{-1.3cm} 
\includegraphics[height=10.0cm,width=04.4cm]{fig_J0238+1637_18nov03_HCT.epsi} 
\hspace*{0.1cm} 
\includegraphics[height=10.0cm,width=04.4cm]{fig_J0423-0120_19nov03_HCT.epsi} 
\hspace*{0.1cm} 
\includegraphics[height=10.0cm,width=04.4cm]{fig_J0423-0120_08dec04_ST.epsi} 
\hspace*{0.1cm} 
\includegraphics[height=10.0cm,width=04.4cm]{fig_J0423-0120_25oct09_ST.epsi} 
} 
\vspace*{50pt} 
\hbox{ 
\hspace*{-1.3cm} 
\includegraphics[height=10.0cm,width=04.4cm]{fig_J0739+0137_05dec05_HCT.epsi} 
\hspace*{0.1cm} 
\includegraphics[height=10.0cm,width=04.4cm]{fig_J0739+0137_06dec05_HCT.epsi} 
\hspace*{0.1cm} 
\includegraphics[height=10.0cm,width=04.4cm]{fig_J0739+0137_09dec05_HCT.epsi} 
\hspace*{0.1cm} 
\includegraphics[height=10.0cm,width=04.4cm]{fig_J1058+0133_25mar07_ST.epsi}  
} 
\caption{ 
The intranight optical DLCs of the HPCDQ monitored in the present study. 
For each night, the source name, the telescope used,  
the date, and the duration of monitoring are given at the top. 
The upper 3 panels show the DLCs of the HPCDQ relative to 3 
comparison stars while the attached lower 3 panels show the star-star DLCs, where 
the solid horizontal lines mark the mean for each star-star DLC. 
The bottom panel gives the plots of seeing variation for the night, 
based on 3 stars monitored along with the blazar on the same CCD frame.  
} 
\label{fig:2} 
\end{figure*} 
 
\newpage 
\begin{figure*} 
\hbox{ 
\hspace*{-1.3cm} 
\includegraphics[height=10.0cm,width=04.4cm]{fig_J1058+0133_16apr07_ST.epsi} 
\hspace*{0.1cm} 
\includegraphics[height=10.0cm,width=04.4cm]{fig_J1058+0133_23apr07_ST.epsi} 
\hspace*{0.1cm} 
\includegraphics[height=10.0cm,width=04.4cm]{fig_J1159+2914_31mar12_IGO.epsi}
\hspace*{0.1cm} 
\includegraphics[height=10.0cm,width=04.4cm]{fig_J1159+2914_01apr12_IGO.epsi}
} 
\vspace*{50pt} 
\hbox{ 
\hspace*{-1.3cm} 
\includegraphics[height=10.0cm,width=04.4cm]{fig_J1159+2914_02apr12_IGO.epsi}
\hspace*{0.1cm}
\includegraphics[height=10.0cm,width=04.4cm]{fig_J1256-0547_26jan06_ST.epsi}
\hspace*{0.1cm}
\includegraphics[height=10.0cm,width=04.4cm]{fig_J1256-0547_28feb06_ST.epsi}
\hspace*{0.1cm}
\includegraphics[height=10.0cm,width=04.4cm]{fig_J1256-0547_20apr09_ST.epsi} 
} 
\begin{center} 
{{\bf Figure~\ref{fig:2}}. \textit {continued}} 
\end{center} 
\end{figure*}  
\clearpage 

\newpage
\begin{figure*}
\hbox{
\hspace*{-1.3cm}
\includegraphics[height=10.0cm,width=04.4cm]{fig_J1512-0906_14jun05_ST.epsi}
\hspace*{0.1cm}
\includegraphics[height=10.0cm,width=04.4cm]{fig_J1512-0906_01may09_ST.epsi}
\hspace*{0.1cm}
\includegraphics[height=10.0cm,width=04.4cm]{fig_J1512-0906_20may09_ST.epsi}
}
\begin{center}
{{\bf Figure~\ref{fig:2}}. \textit {continued}}
\end{center}
\end{figure*}
\clearpage

\newpage 
\begin{figure*}
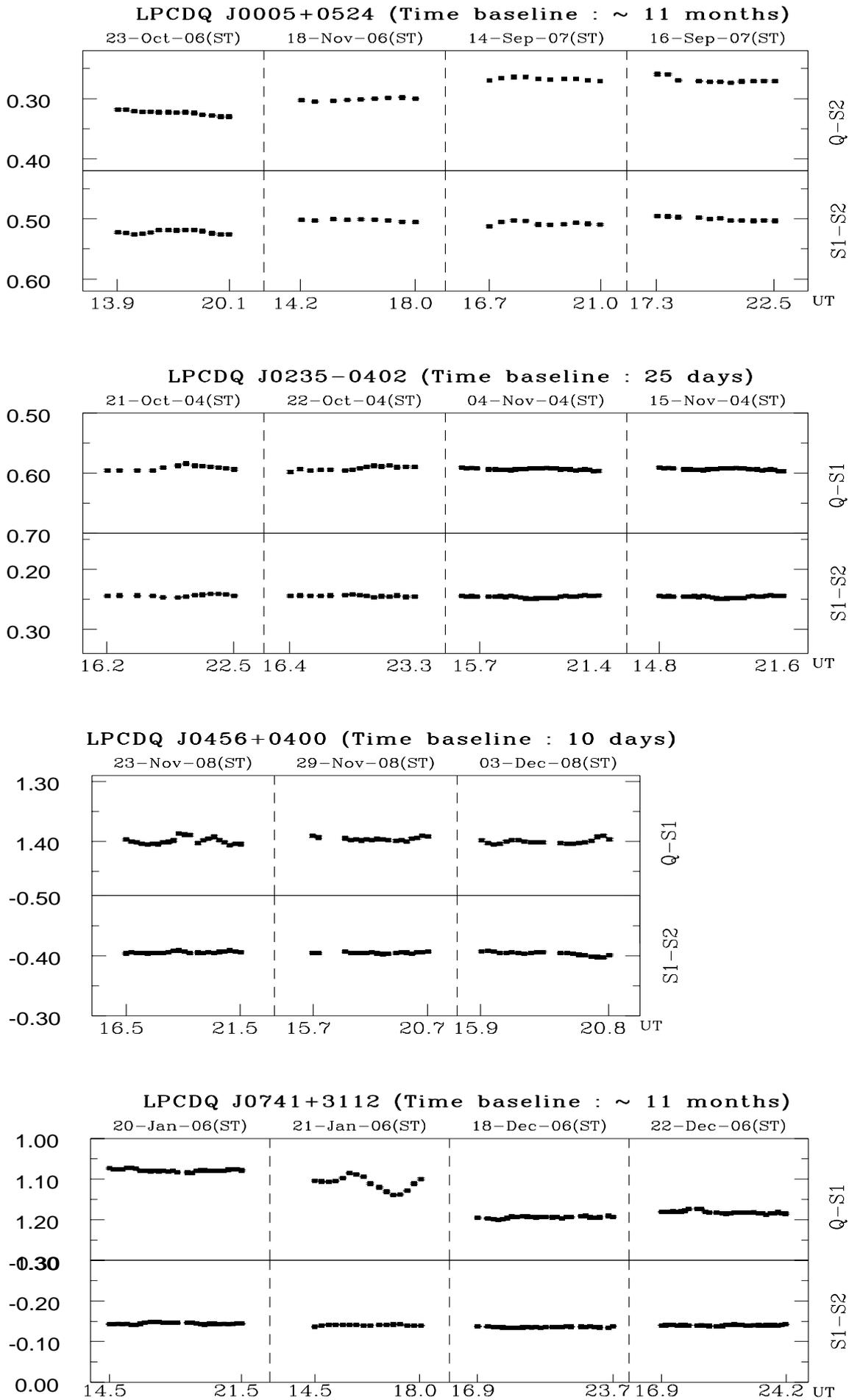
 
\includegraphics[height=05.5cm,width=15.0cm]{ltov_J0005+0524.epsi} 

\vspace*{1.0cm} 
\includegraphics[height=05.5cm,width=15.0cm]{ltov_J0235-0402.epsi} 

\vspace*{1.0cm} 
\includegraphics[height=05.5cm,width=12.0cm]{ltov_J0456+0400.epsi} 

\vspace*{1.0cm} 
\includegraphics[height=05.5cm,width=15.0cm]{ltov_J0741+3112.epsi} 
\caption{Long-term optical variability (LTOV) DLCs for the LPCDQs 
monitored in the present study; source name and the total time span 
covered are  at the top of each panel.  } 
\label{fig:3} 
\end{figure*} 
\clearpage 
 
\newpage 
\begin{figure*} 
\includegraphics[height=05.5cm,width=12.0cm]{ltov_J0842+1835.epsi} 
\vspace*{0.5cm} 

\includegraphics[height=05.5cm,width=12.0cm]{ltov_J1229+0203.epsi} 
\vspace*{0.5cm} 

\includegraphics[height=05.5cm,width=17.0cm]{ltov_J1357+1919.epsi} 
\vspace*{0.5cm} 

\includegraphics[height=05.5cm,width=12.0cm]{ltov_J2203+3145.epsi}  
\begin{center} 
{{\bf Figure~\ref{fig:3}}. \textit {continued}} 
\end{center} 
\end{figure*} 
\clearpage 
 
\newpage 
\begin{figure*}  
\includegraphics[height=05.0cm,width=12.0cm]{ltov_J2346+0930.epsi} 
\begin{center} 
{{\bf Figure~\ref{fig:3}}. \textit {continued}} 
\end{center} 
\end{figure*} 

\newpage 
\begin{figure*}
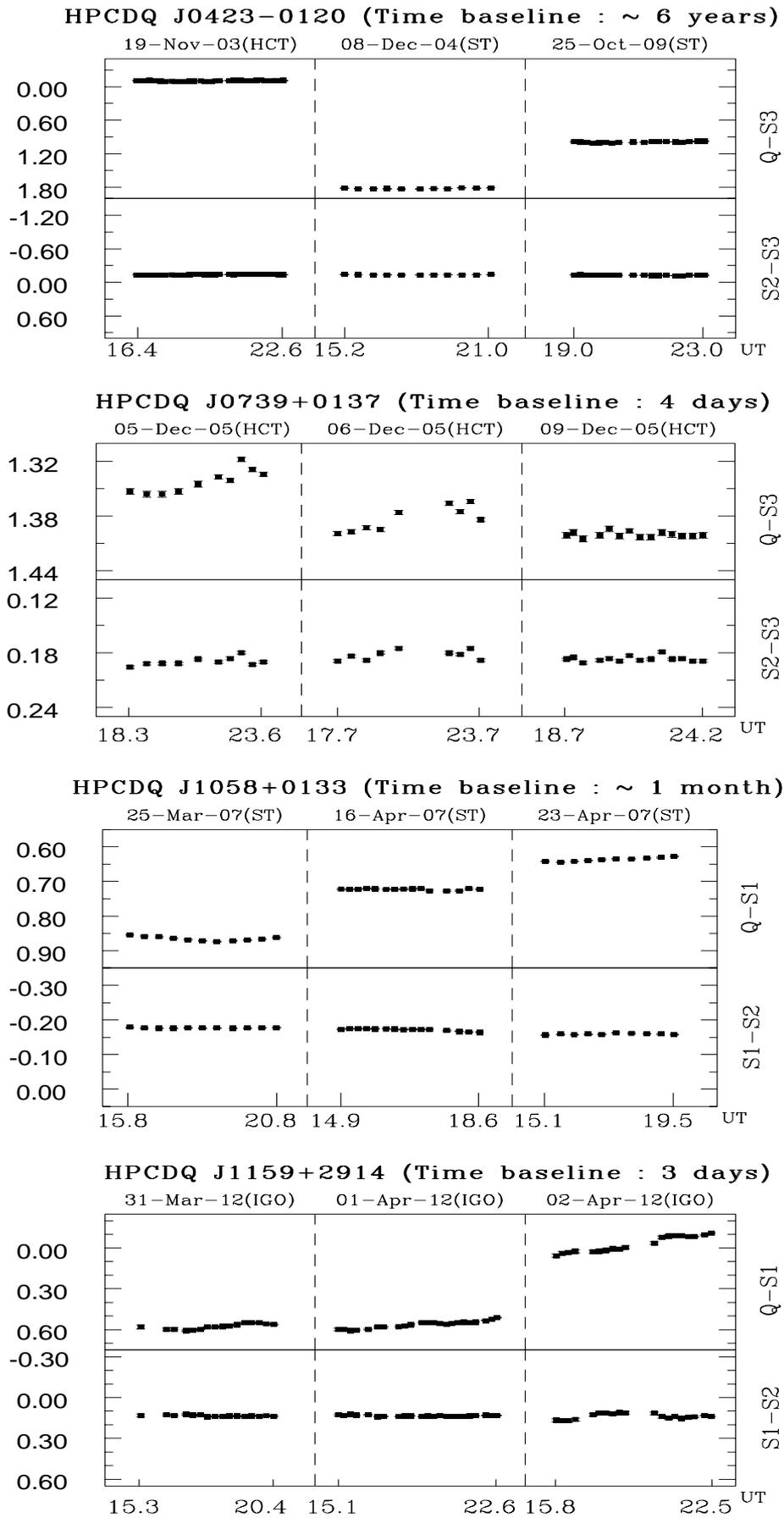
 
\includegraphics[height=05.5cm,width=12.0cm]{ltov_J0423-0120.epsi} 
\vspace*{0.5cm} 

\includegraphics[height=05.5cm,width=12.0cm]{ltov_J0739+0137.epsi} 
\vspace*{0.5cm} 

\includegraphics[height=05.5cm,width=12.0cm]{ltov_J1058+0133.epsi} 
\vspace*{0.5cm} 

\includegraphics[height=05.5cm,width=12.0cm]{ltov_J1159+2914.epsi} 
\caption{As in Fig.~3  for the HPCDQs 
monitored in the present study.  }
\label{fig:4} 
\end{figure*} 
\clearpage 
 
\newpage 
\begin{figure*}  
\includegraphics[height=05.5cm,width=12.0cm]{ltov_J1256-0547.epsi}
\vspace*{0.5cm}

\includegraphics[height=05.0cm,width=12.0cm]{ltov_J1512-0906.epsi} 

\begin{center} 
{{\bf Figure~\ref{fig:4}}. \textit {continued}} 
\end{center} 
\end{figure*} 
\clearpage

\newpage 
\begin{figure*}  
\includegraphics[height=12.0cm,width=12.0cm]{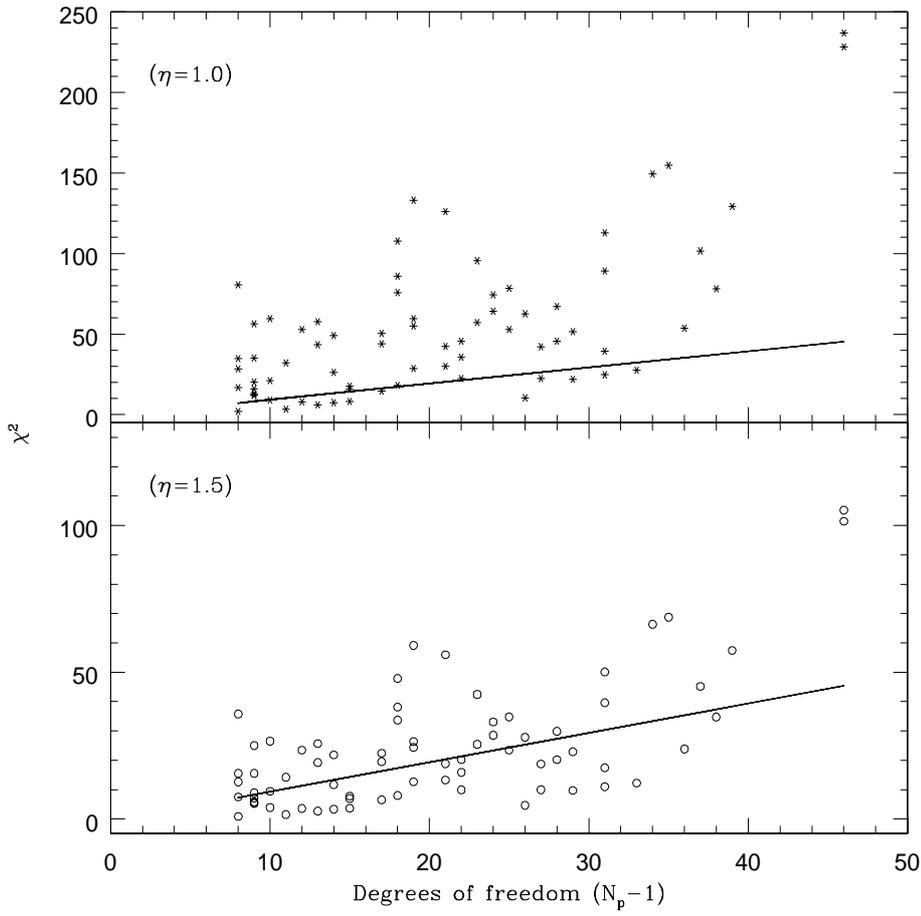} 
\label{fig:5} 
\caption{ Histogram of $\chi^2$ values computed for our entire data set of 73 nights using $\eta = 1.0$ (top)
and $\eta = 1.5$ (bottom). 
The solid line shows the theoretical $\chi^2$ estimator at $p=0.5$ for various degrees of
freedom (see sect. \ref{eta_comp}) . }

\label{etafigure}
\end{figure*} 
\clearpage

\begin{figure}
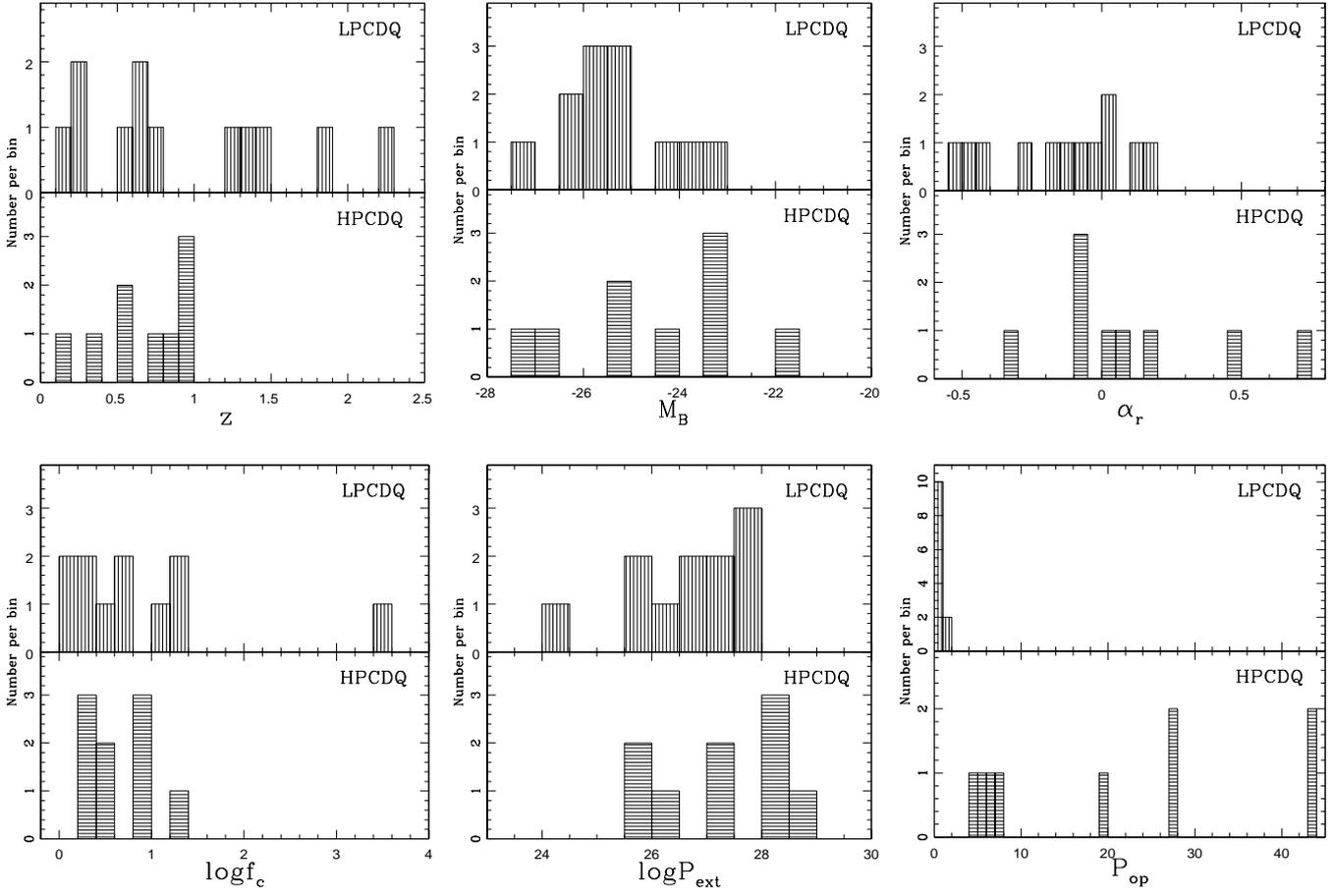
 
\hbox{ 
\hspace*{-1.0cm}\includegraphics[height=  5.7cm, width =  5.7cm, angle = 0]{hist_redshift.epsi} 
\hspace*{0.2cm}\includegraphics[height=  5.7cm, width =  5.7cm, angle = 0]{hist_Mb.epsi} 
\hspace*{0.2cm}\includegraphics[height=  5.7cm, width =  5.7cm, angle = 0]{hist_alpha.epsi} 
} 
\vspace*{0.5cm} 
\hbox{ 
\hspace*{-1.0cm}\includegraphics[height=  5.7cm, width =  5.7cm, angle = 0]{hist_Rcd.epsi} 
\hspace*{0.2cm}\includegraphics[height=  5.7cm, width =  5.7cm, angle = 0]{hist_Pext.epsi} 
\hspace*{0.2cm}\includegraphics[height=  5.7cm, width =  5.7cm, angle = 0]{hist_pop.epsi} 
} 
\caption{Distributions of $z$, $M_B$, $\alpha_r$, $f_{c}$, $P_{ext}$ and $P_{op}$ for
 our two sets of CDQs: LPCDQs (upper panels; vertical stripes); HPCDQs 
(lower panels; horizontal stripes) 
(Sect.\ \ref{discussion}; Table \ref{tab_cdq_parameter}).} 
\medskip 
\label{cdq_dis_properties} 
\end{figure}

\begin{figure} 
\hbox{ 
\hspace*{1.0cm}\includegraphics[height=  8.0cm, width = 12.0cm, angle = 0]{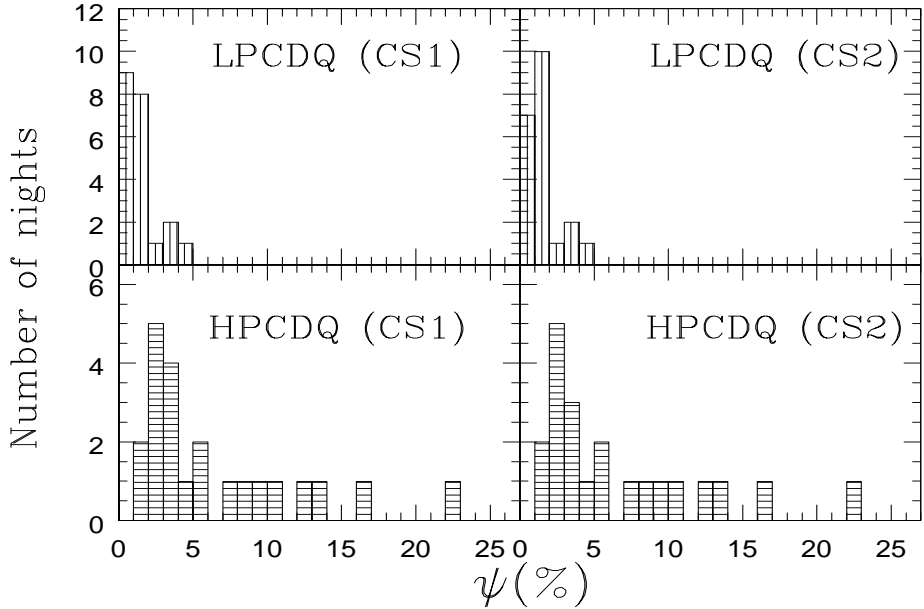} 
} 
\caption{Distribution of INOV amplitude ($\psi$), for LPCDQs (upper panel; 
vertical stripes) and HPCDQs (lower panel; horizontal stripes), estimated from
the DLCs drawn using the two comparison stars, CS1 and CS2.} 
\medskip 
\label{inov_amp_dis} 
\end{figure} 
 
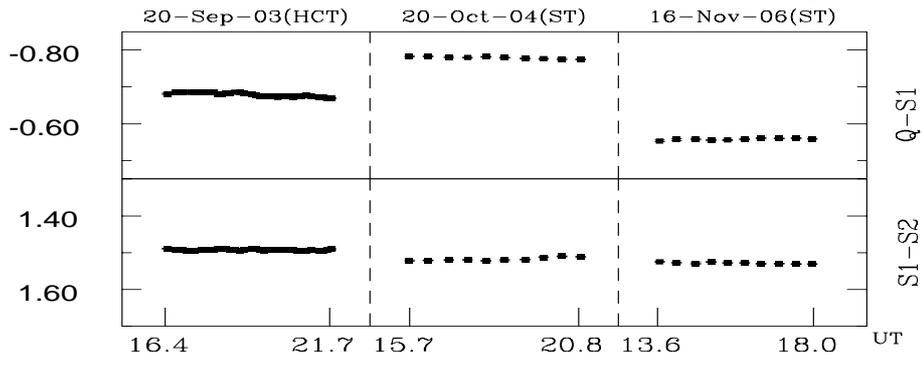
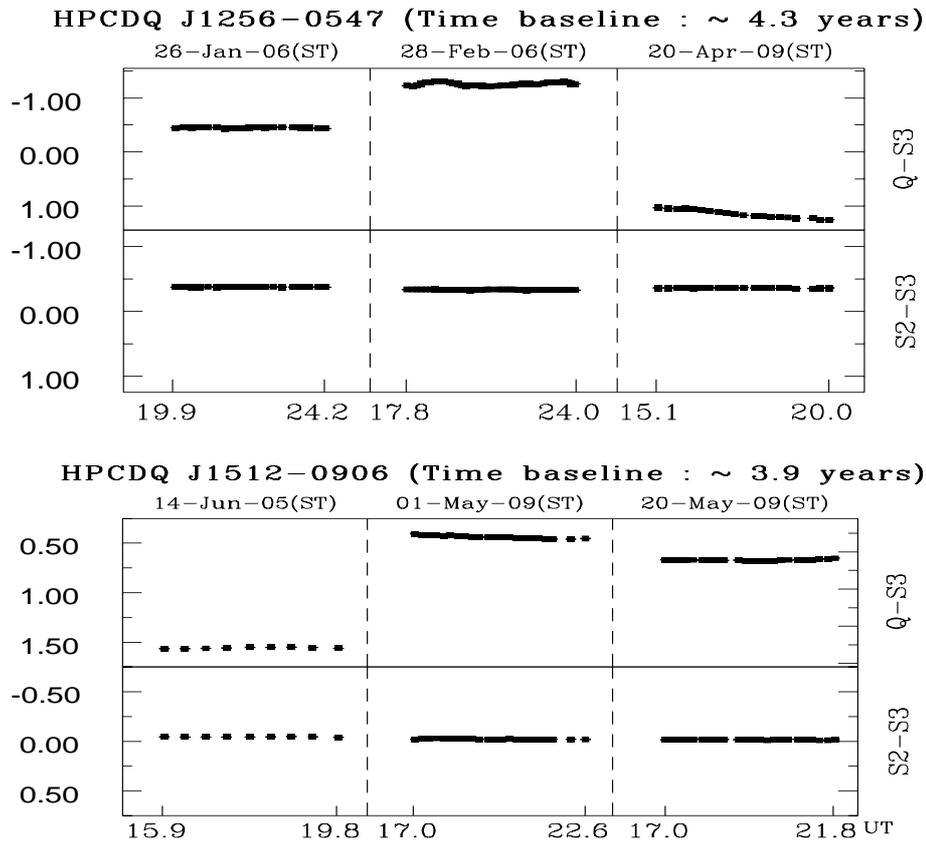
\end{document}